\documentclass[11pt]{article}
\usepackage{amssymb,amsmath,amsfonts}
\usepackage{graphicx}
\usepackage{graphics}
\usepackage{eepic,epsfig}

\@addtoreset{equation}{section}

\makeatother

\begin{document}

\title{Finite temperature fermion condensate, charge and current \\
densities in a (2+1)-dimensional conical space}
\author{S. Bellucci$^{1}$\thanks{%
E-mail: bellucci@lnf.infn.it },\, E. R. Bezerra de Mello$^{2}$\thanks{%
E-mail: emello@fisica.ufpb.br},\, E. Bragan\c{c}a$^{1,2}$\thanks{%
E-mail: deduardo@lnf.infn.it},\, A. A. Saharian$^{3}$ \thanks{%
E-mail: saharian@ysu.am} \\
\\
\textit{$^1$ INFN, Laboratori Nazionali di Frascati,}\\
\textit{Via Enrico Fermi 40, 00044 Frascati, Italy} \vspace{0.3cm}\\
\textit{$^{2}$Departamento de F\'{\i}sica, Universidade Federal da Para\'{\i}%
ba}\\
\textit{58.059-970, Caixa Postal 5.008, Jo\~{a}o Pessoa, PB, Brazil}\vspace{%
0.3cm}\\
\textit{$^3$Department of Physics, Yerevan State University,}\\
\textit{1 Alex Manoogian Street, 0025 Yerevan, Armenia}}
\maketitle

\begin{abstract}
We evaluate the fermion condensate and the expectation values of the charge
and current densities for a massive fermionic field in (2+1)-dimensional
conical spacetime with a magnetic flux located at the cone apex. The
consideration is done for both irreducible representations of the Clifford
algebra. The expectation values are decomposed into the vacuum expectation
values and contributions coming from particles and antiparticles. All these
contributions are periodic functions of the magnetic flux with the period
equal to the flux quantum. Related to the non-invariance of the model under
the parity and time-reversal transformations, the fermion condensate and the
charge density have indefinite parity with respect to the change of the
signs of the magnetic flux and chemical potential. The expectation value of
the radial current density vanishes. The azimuthal current density is the
same for both the irreducible representations of the Clifford algebra. It is
an odd function of the magnetic flux and an even function of the chemical
potential. The behavior of the expectation values in various asymptotic
regions of the parameters are discussed in detail. In particular, we show
that for points near the cone apex the vacuum parts dominate. For a massless
field with zero chemical potential the fermion condensate and charge density
vanish. Simple expressions are derived for the part in the total charge
induced by the planar angle deficit and magnetic flux. Combining the results
for separate irreducible representations, we also consider the fermion
condensate, charge and current densities in parity and time-reversal
symmetric models. Possible applications to graphitic nanocones are discussed.
\end{abstract}

\bigskip


\bigskip

\section{Introduction}

Fermionic field theoretical models in three-dimensional spacetime rise in a
number of physical problems. In particular, the long-wavelength description
of a variety of planar condensed matter systems can be formulated in terms
of the Dirac-like theory. The examples include models of high-temperature
superconductivity, graphene, d-density-wave states and topological
insulators (for reviews see \cite{Frad91,Gusy07}). The long-wavelength
dynamics, equivalent to relativistic Dirac fermions with a controllable
mass, is exhibited by ultracold fermionic atoms in an optical lattice \cite%
{Ruos08}. Another motivation comes from the connection of three-dimensional
models to high temperature behavior in 4-dimensional field theories \cite%
{Meye07}. Field theories in three dimensions also provide simple models in
particle physics. Because of one dimension less they are easier to handle.

Three-dimensional theories exhibit a number of interesting features, such as
flavour symmetry breaking, parity violation, fractionalization of quantum
numbers, that make them interesting on their own. Some special features of
gauge theories, including the supersymmetry breaking, have been discussed in
\cite{Poly77,Dese82}. In three-dimensions, topologically non-trivial gauge
invariant terms in the action provide masses for the gauge fields. The
topological mass term introduces an infrared cutoff in vector gauge theories
providing a way for the solution of the infrared problem without changing
the ultraviolet behavior \cite{Dese82}. If absent at the classical level,
this term is generated through quantum corrections \cite{Niem83}. In the
presence of an external gauge field, fermions in three-dimensional spactime
induce a topologically nontrivial vacuum current having abnormal parity \cite%
{Niem83,Redl84}. In models with fermions coupled to the Chern-Simons gauge
field the Lorentz invariance should be spontaneously broken \cite{Hoso93}:
there exists a state with nonzero magnetic field and with the energy lower
than the lowest energy state in the absence of the magnetic field. Another
interesting feature of the models in two spatial dimensions is the
possibility of the excitations with fractional statistics (anyons) \cite%
{Wilc90}.

The presence of a background gauge field gives rise to the polarization of
the fermionic vacuum. As a consequence, various types of quantum numbers are
generated. In particular, charge and current densities are induced \cite%
{Jaro86}. The vacuum currents induced by cylindrical and toroidal topologies
of background space have been investigated in \cite{Bell09}. Applications
are given to the electronic subsystem of graphene made cylindrical and
toroidal nanotubes, described in terms of the effective Dirac-like theory.
Vacuum expectation values of the current density in locally de Sitter and
anti-de Sitter backgrounds with toroidally compactified spatial dimensions
are studied in \cite{Bell13} and \cite{Beze15}. Among the most interesting
topics in the studies of (2+1)-dimensional theories is the parity and chiral
symmetry-breaking. In particular, it has been shown that a background
magnetic field can serve as a catalyst for the dynamical symmetry breaking
\cite{Gusy94}. A key point of these considerations is the appearance of a
nonzero fermion condensate induced by the magnetic field. This phenomenon
may be important in the physics of high-temperature superconductivity \cite%
{Dore92}.

In the present paper we investigate the finite temperature effects on the
fermion condensate and on the expectation values of the charge and current
densities for a massive fermionic field with nonzero chemical potential in a
(2+1)-dimensional conical spacetime in the presence of a magnetic flux
located at the cone apex. The corresponding vacuum expectation values in the
presence of a circular boundary, concentric with the cone apex, have been
studied in \cite{Beze10}-\cite{Beze12}. The finite temperature effects on
the fermionic condensate and current densities in models with an arbitrary
number of toroidally compact spatial dimensions are discussed in \cite%
{Bell14T}. The finite temperature charge and current densities in the
geometry of a (3+1)-dimensional cosmic string with magnetic flux have been
recently investigated in \cite{Moha15}.

The outline of the paper is as follows. In the next section we consider the
fermion condensate for a two-component spinor field realizing the
irreducible representation of the Clifford algebra. Expressions are derived
for the separate contributions from particles and antiparticles for general
values of the planar angle deficit and for the magnetic flux. Various
asymptotic regions of the parameters are considered in detail. The
expectation values of the charge and current densities are investigated in
Sections \ref{sec:Charge} and \ref{sec:Current}, respectively. We provide
simple expressions for the topological part in the total charge, induced by
the planar angle deficit and by the magnetic flux. The expectation values in
the model with parity and time-reversal invariance are discussed in Section %
\ref{sec:Tsym}. These expectation values are obtained by combining the
results from the previous section for two irreducible representations of the
Clifford algebra. The main results are summarized in Section \ref{sec:Conc}.
In the Appendix we derive the relations used in the evaluation of the
topological part in the total charge.

\section{Fermion condensate}

\label{sec:FC}

We consider a fermionic field $\psi (x)$ on background of (2+1)-dimensional
spacetime assuming that the field is in thermal equilibrium at temperature $%
T $. In this section we evaluate the fermion condensate (FC) defined as the
expectation value $\langle \bar{\psi}\psi \rangle =\mathrm{tr\,}[\hat{\rho}%
\bar{\psi}\psi ]$, with $\bar{\psi}=\psi ^{\dag }\gamma ^{0}$ being the
Dirac adjoint and the angular brackets denote the ensemble average with the
density matrix $\hat{\rho}=Z^{-1}e^{\beta (\hat{H}-\mu ^{\prime }\hat{Q})}$,
where $\beta =1/T$. Here $\hat{H}$ is the Hamilton operator, $\hat{Q}$ is a
conserved charge with the related chemical potential $\mu ^{\prime }$ and $Z=%
\mathrm{tr\,}[e^{-\beta (\hat{H}-\mu ^{\prime }\hat{Q})}]$. The FC is among
the most important characteristic for the system under consideration. In
particular, it plays an important role in the models of dynamical breaking
of chiral symmetry.

The background geometry is described by the (2+1)-dimensional line element
\begin{equation}
ds^{2}=g_{\mu \nu }dx^{\mu }dx^{\nu }=dt^{2}-dr^{2}-r^{2}d\phi ^{2},
\label{linel}
\end{equation}%
where $0\leqslant \phi \leqslant \phi _{0}$. For $\phi _{0}<2\pi $, this
line element describes (2+1)-dimensional conical spacetime with the planar
angle deficit $2\pi -\phi _{0}$. We will discuss the case of two-components
spinor field realizing the irreducible representation of the Clifford
algebra. Assuming the presence of the external electromagnetic field with
the vector potential $A_{\mu }$, the field operator obeys the Dirac equation%
\begin{equation}
\left( i\gamma ^{\mu }D_{\mu }-sm\right) \psi =0,\ D_{\mu }=\partial _{\mu
}+\Gamma _{\mu }+ieA_{\mu },  \label{Dirac}
\end{equation}%
where $\Gamma _{\mu }$ is the spin connection and $e$ is the charge of the
field quantum. Here, $s=+1$ and $s=-1$ correspond to inequivalent
irreducible representations of the Clifford algebra in (2+1)-dimensions (see
Section \ref{sec:Tsym}). With these representations, the mass term violates
the parity and time-reversal invariances. In the coordinates corresponding
to (\ref{linel}), the gamma matrices can be taken in the representation%
\begin{equation}
\gamma ^{0}=\left(
\begin{array}{cc}
1 & 0 \\
0 & -1%
\end{array}%
\right) ,\quad \gamma ^{l}=\frac{i^{2-l}}{r^{l-1}}\left(
\begin{array}{cc}
0 & e^{-iq\phi } \\
(-1)^{l-1}e^{iq\phi } & 0%
\end{array}%
\right) ,  \label{gamma}
\end{equation}%
with $l=1,2$ and $q=2\pi /\phi _{0}$.

In what follows, in the region $r>0$, we consider the gauge field
configuration $A_{\mu }=(0,0,A)$, where $A_{2}=A$ is the covariant component
of the vector potential in the coordinates $(t,r,\phi )$. For the
corresponding physical component one has $A_{\phi }=-A/r$. This corresponds
to an infinitely thin magnetic flux $\Phi =-\phi _{0}A$ located at $r=0$. As
it will be seen below, in the expressions for the expectation values the
parameter $A$ enters in the form of the combination%
\begin{equation*}
\alpha =eA/q=-e\Phi /(2\pi ),
\end{equation*}%
where $q=2\pi /\phi _{0}$. We decompose it as
\begin{equation}
\alpha =\alpha _{0}+n_{0},\quad |\alpha _{0}|<1/2,  \label{alpha}
\end{equation}%
with $n_{0}$ being an integer. It is the fractional part $\alpha _{0}$ which
is responsible for physical effects.

Let us denote by $\psi _{\sigma }^{(+)}(x)$ and $\psi _{\sigma }^{(-)}(x)$ a
complete orthonormal set of the positive- and negative-energy solutions of
the equation (\ref{Dirac}). These solutions are labeled by a set of quantum
numbers $\sigma $. Expanding the field operator $\psi (x)$ in terms of the
functions $\psi _{\sigma }^{(\pm )}(x)$ and using the commutation relations
for the annihilation and creation operators, the following decomposition is
obtained for the FC:%
\begin{equation}
\langle \bar{\psi}\psi \rangle =\langle \bar{\psi}\psi \rangle _{0}+\langle
\bar{\psi}\psi \rangle _{+}+\langle \bar{\psi}\psi \rangle _{-}.
\label{FCdec}
\end{equation}%
Here,
\begin{equation}
\langle \bar{\psi}\psi \rangle _{0}=\sum_{\sigma }\bar{\psi}_{\sigma
}^{(-)}(x)\psi _{\sigma }^{(-)}(x),  \label{FCvac}
\end{equation}%
is the FC in the vacuum state and $\langle \bar{\psi}\psi \rangle _{+}$ and $%
\langle \bar{\psi}\psi \rangle _{-}$ are the contributions from particles
and antiparticles, respectively. The latter are given by the expressions%
\begin{equation}
\langle \bar{\psi}\psi \rangle _{\pm }=\pm \sum_{\sigma }\frac{\bar{\psi}%
_{\sigma }^{(\pm )}(x)\psi _{\sigma }^{(\pm )}(x)}{e^{\beta (E_{\sigma }\mp
\mu )}+1},  \label{FCpm}
\end{equation}%
where $\mu =e\mu ^{\prime }$ and $\pm E_{\sigma }$ are the energies
corresponding to the modes $\psi _{\sigma }^{(\pm )}(x)$. In (\ref{FCvac})
and (\ref{FCpm}), $\sum_{\sigma }$ includes the summation over the discrete
quantum numbers and the integration over the continuous ones. The modes are
normalized in accordance with the standard normalization condition $\int
d^{2}x\,\sqrt{\gamma }\psi _{\sigma }^{(\pm )\dag }(x)\psi _{\sigma ^{\prime
}}^{(\pm )}(x)=\delta _{\sigma \sigma ^{\prime }}$, with $\gamma $ being the
determinant of the spatial metric tensor. The part in the FC corresponding
to the vacuum expectation value, $\langle \bar{\psi}\psi \rangle _{0}$, has
been investigated in \cite{Bell11} and here we will be mainly concerned with
the finite temperature effects.

For the evaluation of the FC in accordance with (\ref{FCpm}) we need to know
the mode functions. The general solution of the radial equation for these
functions contains the part regular at the origin and the part which
diverges at $r=0$. Under the condition%
\begin{equation}
2|\alpha _{0}|\leqslant 1-1/q,  \label{condalf0}
\end{equation}%
the irregular modes are eliminated by the normalizability condition. In the
case $2|\alpha _{0}|>1-1/q$, there are irregular normalizable modes and for
the unique identification of the functions an additional boundary condition
is required at the origin. In the literature it has been already shown that
the standard procedure for the self-adjoint extension of the Dirac
Hamiltonian gives rise to a one-parameter family of boundary conditions in
the background of an Aharonov-Bohm gauge field \cite{Sous89}. The specific
value of the parameter is related to the physical details of the magnetic
field distribution inside a more realistic finite radius flux tube (for a
more detailed discussion and for specific models with finite radius magnetic
flux see \cite{Hage90}). The idealized model under consideration is a
limiting case of the latter.

Following \cite{Beze10}, for irregular modes we consider a special case of
boundary conditions at the cone apex, when the bag boundary condition is
imposed at a finite radius, which is then taken to zero. The bag boundary
condition ensures the zero flux of fermions and this corresponds to an
impenetrable flux tube. The corresponding mode functions have the form%
\begin{equation}
\psi _{\sigma }^{(\pm )}(x)=c_{0}^{(\pm )}e^{iqj\phi \mp iEt}\left(
\begin{array}{c}
J_{\beta _{j}}(\gamma r)e^{-iq\phi /2} \\
\epsilon _{j}\frac{\gamma e^{iq\phi /2}}{\pm E+sm}J_{\beta _{j}+\epsilon
_{j}}(\gamma r)%
\end{array}%
\right) ,  \label{Modes}
\end{equation}%
where $j=\pm 1/2,\pm 3/2,\ldots $, $J_{\nu }(x)$ is the Bessel function of
the first kind and%
\begin{equation}
\beta _{j}=q|j+\alpha |-\epsilon _{j}/2,  \label{betj}
\end{equation}%
with $\epsilon _{j}=1$ for $j>-\alpha $ and $\epsilon _{j}=-1$ for $%
j<-\alpha $. The spinors (\ref{Modes}) are specified by the set $\sigma
=(\gamma ,j)$ with $0\leqslant \gamma <\infty $ and for the energy one has $%
E=E_{\sigma }=\sqrt{\gamma ^{2}+m^{2}}$. The normalization coefficients are
given by the expression%
\begin{equation}
c_{0}^{(\pm )2}=\gamma \frac{E\pm sm}{2\phi _{0}E}.  \label{cpm}
\end{equation}%
The modes (\ref{Modes}) are eigenfunctions of the angular momentum operator $%
\hat{J}=-(i/q)(\partial _{\phi }+ieA)+\sigma _{3}/2$, $\sigma _{3}=\mathrm{%
diag}(1,-1)$, for the eingenvalues $j+\alpha $:%
\begin{equation}
\hat{J}\psi _{\sigma }^{(\pm )}(x)=(j+\alpha )\psi _{\sigma }^{(\pm )}(x).
\label{J}
\end{equation}%
In the mode-sum formulas (\ref{FCvac}) and (\ref{FCpm}) one has $%
\sum_{\sigma }=\sum_{j}\int_{0}^{\infty }d\gamma $, with the summation over $%
j=\pm 1/2,\pm 3/2,...$. The parameter $\alpha $ is the magnetic flux
measured in units of the flux quantum. In the case when the parameter $%
\alpha $ is half of an odd integer the mode with $j=-\alpha $ should be
considered separately. The corresponding mode function is presented in \cite%
{Beze10}. In order do not complicate the consideration, in the following
discussion we will exclude this case. Note that, under the condition $%
2|\alpha _{0}|>1-1/q$, the irregular mode corresponds to $j=-n_{0}-\mathrm{%
sgn}(\alpha _{0})/2$. The expectation values for general case of boundary
condition at the cone apex are considered in a way similar to that described
below. The corresponding results differ by the contribution of the irregular
mode only.

Using the mode functions (\ref{Modes}), the contributions to the FC\ from
particles and antiparticles are presented in the form%
\begin{equation}
\langle \bar{\psi}\psi \rangle _{\pm }=\frac{q}{4\pi }\sum_{j}\int_{0}^{%
\infty }d\gamma \frac{\gamma }{e^{\beta (E\mp \mu )}+1}\left\{ \frac{sm}{E}%
[J_{\beta _{j}}^{2}(\gamma r)+J_{\beta _{j}+\epsilon _{j}}^{2}(\gamma r)]\pm
\lbrack J_{\beta _{j}}^{2}(\gamma r)-J_{\beta _{j}+\epsilon _{j}}^{2}(\gamma
r)]\right\} .  \label{FCpm1}
\end{equation}%
For the further transformation of this expression we use the integral
representation of the Bessel function \cite{Wats66}:%
\begin{equation}
J_{\nu }^{2}(x)=\frac{1}{2\pi i}\int_{c-i\infty }^{c+i\infty }\frac{dt}{t}%
e^{t/2-x^{2}/t}I_{\nu }(x^{2}/t),  \label{IntJ}
\end{equation}%
where $c$ is a positive constant and $I_{\nu }(y)$ is the modified Bessel
function of the first kind. Inserting into (\ref{FCpm1}), the separate
contributions $\langle \bar{\psi}\psi \rangle _{\pm }$ are expressed in
terms of the series
\begin{equation}
\mathcal{I}(q,\alpha ,z)=\sum_{j}I_{\beta _{j}}(z).  \label{series}
\end{equation}%
The function $\mathcal{I}(q,\alpha ,z)$ is periodic with respect to $\alpha $
with the period 1. Hence, if we present the parameter $\alpha $ in the form (%
\ref{alpha}), then $\mathcal{I}(q,\alpha ,z)$ does not depend on $n_{0}$ and
$\mathcal{I}(q,\alpha ,z)=\mathcal{I}(q,\alpha _{0},z)$. For the second
series, appearing after the application of (\ref{IntJ}), one has $%
\sum_{j}I_{\beta _{j}+\epsilon _{j}}(z)=\mathcal{I}(q,-\alpha _{0},z)$. As a
result, the expression on the right of (\ref{FCpm1}) is transformed as%
\begin{equation}
\langle \bar{\psi}\psi \rangle _{\pm }=\frac{q}{4\pi }\int_{0}^{\infty
}d\gamma \frac{\gamma }{e^{\beta (E\mp \mu )}+1}\frac{1}{2\pi i}%
\int_{c-i\infty }^{c+i\infty }\frac{dt}{t}e^{t/2-\gamma
^{2}r^{2}/t}\sum_{\delta =\pm 1}(sm/E\pm \delta )\mathcal{I}(q,\delta \alpha
_{0},\gamma ^{2}r^{2}/t).  \label{FCpm11}
\end{equation}

For the series (\ref{series}) one has the representation \cite{Beze10}%
\begin{eqnarray}
\mathcal{I}(q,\alpha _{0},x) &=&\frac{2}{q}\sideset{}{'}{\sum}%
_{l=0}^{p}(-1)^{l}\cos (\pi l(2\alpha _{0}-1/q))e^{x\cos (2\pi l/q)}  \notag
\\
&&-\frac{1}{\pi }\int_{0}^{\infty }dy\frac{e^{-x\cosh y}f(q,\alpha _{0},y)}{%
\cosh (qy)-\cos (q\pi )},  \label{Ic1}
\end{eqnarray}%
where $p$ is the integer part of $q/2$, $p=[q/2]$, and
\begin{equation}
f(q,\alpha _{0},y)=\sum_{\delta =\pm 1}\delta \cos (q\pi (1/2-\delta \alpha
_{0}))\cosh ((q\alpha _{0}+\delta q/2-1/2)y).  \label{fq1}
\end{equation}%
The prime on the sign of the summation in (\ref{Ic1}) means that the term $%
l=0$ and the term $l=q/2=p$ for even values of $q$ should be taken with the
coefficient 1/2. In the range $1\leqslant q<2$, the sum over $l$ in (\ref%
{Ic1}) is absent. Note that for $q$ being an even integer, $q=2p$, for the
function (\ref{fq1}) one has%
\begin{equation}
f(q,\alpha _{0},y)=2(-1)^{p}\cos (\pi q\alpha _{0})\sinh ((q\alpha
_{0}-1/2)y)\sinh (py),  \label{fq12}
\end{equation}%
and the integrand in (\ref{Ic1}) is regular in the lower limit of
integration.

Substituting the representation (\ref{Ic1}) into (\ref{FCpm11}), the $t$%
-integral is evaluated with the help of the formula \cite{Wats66}%
\begin{equation}
\frac{1}{2\pi i}\int_{c-i\infty }^{c+i\infty }\frac{dt}{t}%
e^{t/2-2b^{2}/t}=J_{0}(2b).  \label{IntJ0}
\end{equation}%
As a result, the following expression is obtained%
\begin{eqnarray}
\langle \bar{\psi}\psi \rangle _{\pm } &=&\langle \bar{\psi}\psi \rangle
_{\pm }^{(M)}+\frac{1}{\pi }\int_{0}^{\infty }d\gamma \frac{\gamma }{%
e^{\beta (E\mp \mu )}+1}  \notag \\
&&\times \left\{ \sum_{l=1}^{p}(-1)^{l}\left[ \frac{sm}{E}c_{l}\cos (2\pi
l\alpha _{0})\pm s_{l}\sin (2\pi l\alpha _{0})\right] J_{0}(2\gamma
rs_{l})\right.  \notag \\
&&\left. -\frac{q}{\pi }\int_{0}^{\infty }dy\frac{\frac{sm}{E}f_{1}(q,\alpha
_{0},y)\pm f_{2}(q,\alpha _{0},y)}{\cosh (2qy)-\cos (q\pi )}J_{0}(2\gamma
r\cosh y)\right\} ,  \label{FCpm12}
\end{eqnarray}%
with the notations%
\begin{equation}
c_{l}=\cos (\pi l/q),\;s_{l}=\sin (\pi l/q),  \label{clsl}
\end{equation}%
and%
\begin{eqnarray}
f_{1}(q,\alpha _{0},y) &=&-\sinh y\sum_{\delta =\pm 1}\cos (q\pi (1/2-\delta
\alpha _{0}))\sinh ((1+2\delta \alpha _{0})qy),  \notag \\
f_{2}(q,\alpha _{0},y) &=&\cosh y\sum_{\delta =\pm 1}\delta \cos (q\pi
(1/2-\delta \alpha _{0}))\cosh ((1+2\delta \alpha _{0})qy).  \label{f2q}
\end{eqnarray}%
Note that we have $f_{l}(q,\alpha _{0},y)=\sum_{n=\pm 1}\delta
^{n-1}f(q,n\alpha _{0},2y)/2$, $n=1,2$. In (\ref{FCpm12}), the part
\begin{equation}
\langle \bar{\psi}\psi \rangle _{\pm }^{(M)}=\frac{sm}{2\pi }%
\int_{0}^{\infty }d\gamma \frac{\gamma /E}{e^{\beta (E\mp \mu )}+1},
\label{FCpm1M}
\end{equation}%
is the contribution coming from the $l=0$ term in the right-hand side of (%
\ref{Ic1}) and it coincides with the corresponding quantity in
(2+1)-dimensional Minkowski spacetime ($q=1$) in the absence of the magnetic
flux ($\alpha _{0}=0$).

For an integer $q$, the expression (\ref{FCpm12}) is essentially simplified
for special values of $\alpha _{0}$ defined by%
\begin{equation}
\alpha _{0}=\frac{n+1/2-\{q/2\}}{q},  \label{alfsp}
\end{equation}%
where $n$ is an integer and the figure braces stand for the fractional part
of the enclosed expression. The allowed values for $n$ are obtained from the
condition $|\alpha _{0}|<1/2$. For $\alpha _{0}$ from (\ref{alfsp}) one has $%
\cos (q\pi (1/2-\delta \alpha _{0}))=0$ and, hence, both the functions $%
f_{l}(q,\alpha _{0},y)$ in (\ref{f2q}) become zero. As a result, the
integral term in the figure braces of (\ref{FCpm12}) vanishes.

Returning to the general values of the parameters $q$ and $\alpha _{0}$,
first we will consider the case with $|\mu |\leqslant m$. By using the
relation%
\begin{equation}
\left( e^{y}+1\right) ^{-1}=-\sum_{n=1}^{\infty }(-1)^{n}e^{-ny},
\label{Expansion}
\end{equation}%
after the evaluation of the $\gamma $-integrals, the following
representation is obtained%
\begin{eqnarray}
\langle \bar{\psi}\psi \rangle _{\pm } &=&\langle \bar{\psi}\psi \rangle
_{\pm }^{(M)}-\frac{4m^{2}}{(2\pi )^{3/2}}\sum_{n=1}^{\infty }(-1)^{n}e^{\pm
n\beta \mu }  \notag \\
&&\times \left\{ \sum_{l=1}^{p}(-1)^{l}\left[ sc_{l}\cos (2\pi l\alpha
_{0})f_{1/2}(c_{nl})\pm mn\beta s_{l}\sin (2\pi l\alpha _{0})f_{3/2}(c_{nl})%
\right] \right.  \notag \\
&&\left. -\frac{q}{\pi }\int_{0}^{\infty }dy\left[ \frac{sf_{1}(q,\alpha
_{0},y)f_{1/2}(c_{n}(y))}{\cosh (2qy)-\cos (q\pi )}\pm mn\beta \frac{%
f_{2}(q,\alpha _{0},y)f_{3/2}(c_{n}(y))}{\cosh (2qy)-\cos (q\pi )}\right]
\right\} ,  \label{FCpm5}
\end{eqnarray}%
where $f_{\nu }(x)=K_{\nu }(x)/x^{\nu }$, with $K_{\nu }(x)$ being the
Macdonald function, and%
\begin{equation}
\langle \bar{\psi}\psi \rangle _{\pm }^{(M)}=-\frac{sm}{2\pi \beta }%
\sum_{n=1}^{\infty }\frac{(-1)^{n}}{n}e^{\left( \pm \mu -m\right) n\beta }=%
\frac{sm}{2\pi \beta }\ln \left( 1+e^{\left( \pm \mu -m\right) \beta
}\right) .  \label{FCpmM}
\end{equation}%
In (\ref{FCpm5}), we have introduced the notations%
\begin{eqnarray}
c_{nl} &=&m\sqrt{n^{2}\beta ^{2}+4r^{2}\sin ^{2}(\pi l/q)},  \notag \\
c_{n}(y) &=&m\sqrt{n^{2}\beta ^{2}+4r^{2}\cosh ^{2}y}.  \label{cny}
\end{eqnarray}%
For the functions $f_{\nu }(x)$ in (\ref{FCpm5}) we have%
\begin{equation}
f_{1/2}(x)=\sqrt{\frac{\pi }{2}}\frac{e^{-x}}{x},\;f_{3/2}(x)=\sqrt{\frac{%
\pi }{2}}\frac{e^{-x}}{x^{3}}(1+x).  \label{fs}
\end{equation}

Now summing all the contributions, for the FC we get%
\begin{eqnarray}
\langle \bar{\psi}\psi \rangle &=&\langle \bar{\psi}\psi \rangle
_{0}+\langle \bar{\psi}\psi \rangle ^{(M)}-\frac{2^{3/2}m^{2}}{\pi ^{3/2}}%
\sum_{n=1}^{\infty }(-1)^{n}\left\{ \cosh (n\beta \mu )\left[
\sum_{l=1}^{p}(-1)^{l}\right. \right.  \notag \\
&&\times \left. sc_{l}\cos (2\pi l\alpha _{0})f_{1/2}(c_{nl})-\frac{sq}{\pi }%
\int_{0}^{\infty }dy\,\frac{f_{1}(q,\alpha _{0},y)f_{1/2}(c_{n}(y))}{\cosh
(2qy)-\cos (q\pi )}\right]  \notag \\
&&+mn\beta \sinh (n\beta \mu )\left[ \sum_{l=1}^{p}(-1)^{l}s_{l}\sin (2\pi
l\alpha _{0})f_{3/2}(c_{nl})\right.  \notag \\
&&\left. \left. -\frac{q}{\pi }\int_{0}^{\infty }dy\,\frac{f_{2}(q,\alpha
_{0},y)f_{3/2}(c_{n}(y))}{\cosh (2qy)-\cos (q\pi )}\right] \right\} ,
\label{FC}
\end{eqnarray}%
with the Minkowskian part%
\begin{equation}
\langle \bar{\psi}\psi \rangle ^{(M)}=\frac{sm}{2\pi \beta }\left[ \ln
(1+e^{-\beta (m-\mu )})+\ln (1+e^{-\beta (m+\mu )})\right] .  \label{FCM}
\end{equation}%
The expression for the FC in the vacuum state can be found in \cite{Bell11}:
\begin{eqnarray}
&&\langle \bar{\psi}\psi \rangle _{0}=-\frac{sm}{2\pi r}\Big[%
\sum_{l=1}^{p}(-1)^{l}\frac{c_{l}}{s_{l}}\cos (2\pi l\alpha
_{0})e^{-2mrs_{l}}  \notag \\
&&-\frac{q}{\pi }\int_{0}^{\infty }dy\frac{f_{1}(q,\alpha _{0},y)}{\cosh
(2qy)-\cos (q\pi )}\frac{e^{-2mr\cosh y}}{\cosh y}\Big].  \label{FC0}
\end{eqnarray}%
This contribution can be combined with the sum over $n$ in (\ref{FC})
containing the factor $\cosh (n\beta \mu )$ writing $\sum_{n=0}^{\infty
\prime }$ instead of $\sum_{n=1}^{\infty }$. The prime on the summation sign
means that the term $n=0$ should be taken with the coefficient 1/2. The FC
is a periodic function of the magnetic flux with the period equal to the
flux quantum. Note that it has no definite parity with respect to the
reflections $\alpha _{0}\rightarrow -\alpha _{0}$ and $\mu \rightarrow -\mu $%
. This is related to the non-invariance of the model under the parity ($P$-)
and time-reversal ($T$-) transformations (see the discussion in Section \ref%
{sec:Tsym}). Note that the FC in the vacuum state is an even function of $%
\alpha _{0}$.

The general formula (\ref{FC}) is further simplified in various special
cases. For a massless fermionic field, because of the condition $|\mu
|\leqslant m$, we should also assume that $\mu =0$ and the FC vanishes. In
the special case when the magnetic flux is absent, one has $\alpha _{0}=0$,
and the equation (\ref{FC}) becomes%
\begin{eqnarray}
\langle \bar{\psi}\psi \rangle &=&\langle \bar{\psi}\psi \rangle ^{(M)}-%
\frac{2sm^{2}}{\pi }\sideset{}{'}{\sum}_{n=0}^{\infty }(-1)^{n}\cosh {%
(n\beta \mu )}\left[ \sum_{l=1}^{p}(-1)^{l}c_{l}\frac{e^{-c_{nl}}}{c_{nl}}%
\right.  \notag \\
&&+\left. \frac{2q}{\pi }\cos {(q\pi /2)}\int_{0}^{\infty }dy\,\frac{\sinh {%
(2qy)}\sinh {y}}{\cosh (2qy)-\cos (q\pi )}\frac{e^{-c_{n}(y)}}{c_{n}(y)}%
\right] .  \label{FC12}
\end{eqnarray}%
In this case the FC is an even function of the chemical potential. For the
background of Minkowski spacetime with the magnetic flux we take $q=1$ and
the FC is given by%
\begin{eqnarray}
\langle \bar{\psi}\psi \rangle &=&\langle \bar{\psi}\psi \rangle ^{(M)}+%
\frac{2^{3/2}m^{2}}{\pi ^{5/2}}\sin (\pi \alpha _{0})\sideset{}{'}{\sum}%
_{n=0}^{\infty }(-1)^{n}\int_{0}^{\infty }dy\,  \notag \\
&&\times \left[ -s\cosh (n\beta \mu )\tanh y\sinh (2\alpha
_{0}y)f_{1/2}(c_{n}(y))\right.  \notag \\
&&\left. +mn\beta \sinh (n\beta \mu )\cosh (2\alpha _{0}y)f_{3/2}(c_{n}(y))
\right] .  \label{FCq1}
\end{eqnarray}%
The $n=0$ terms in (\ref{FC12}) and (\ref{FCq1}) correspond to the vacuum
expectation values.

Let us consider the asymptotic behavior of the FC in the limiting regions of
the parameters. For points near the apex of the cone two separate regions of
the values of the parameter $\alpha _{0}$ should be distinguished. For $%
2|\alpha _{0}|<1-1/q$, the thermal part is finite on the apex and the
corresponding expression is directly obtained from (\ref{FC}) putting $r=0$.
Recall that, in this case, all the modes (\ref{Modes}) are regular at the
apex. For $2|\alpha _{0}|>1-1/q$ the integrals diverge on the apex. In order
to find the leading term in the asymptotic expansion over the distance from
the origin, we note that for points near the apex the dominant contribution
to the integrals in (\ref{FC}) comes from large values of $y$. Expanding the
integrands, it can be seen that%
\begin{eqnarray}
\langle \bar{\psi}\psi \rangle &\approx &\langle \bar{\psi}\psi \rangle _{0}+%
\frac{m^{2}q\Gamma (1/2-\rho )}{2^{\rho }\pi ^{5/2}\left( mr\right)
^{1-2\rho }}\cos \left( \pi \rho \right) \sum_{n=1}^{\infty }(-1)^{n}  \notag
\\
&&\times \left[ \mathrm{sgn}(\alpha _{0})mn\beta \sinh (n\beta \mu )f_{\rho
+1}(mn\beta )-s\cosh (n\beta \mu )f_{\rho }(mn\beta )\right] ,
\label{FCnear0}
\end{eqnarray}%
with the notation $\rho =q\left( 1/2-|\alpha _{0}|\right) $. Note that in
the case under consideration one has $\rho <1/2$. Hence, for $2|\alpha
_{0}|>1-1/q$ the finite temperature part diverges on the apex as $r^{2\rho
-1}$. The vacuum expectation value $\langle \bar{\psi}\psi \rangle _{0}$
behaves as $1/r$ and it dominates for small $r$.

Now we turn to the investigation of the asymptotics for the FC at low and
high temperatures. In the low temperature limit the main contribution to the
thermal part comes from the $n=1$ term and to the leading order we find%
\begin{eqnarray}
\langle \bar{\psi}\psi \rangle &\approx &\langle \bar{\psi}\psi \rangle _{0}+%
\frac{m}{\pi \beta }e^{-\beta (m-|\mu |)}\left[ \sideset{}{'}{\sum}%
_{l=0}^{p}(-1)^{l}\cos (\pi l(2s\alpha _{0}\mathrm{sgn}(\mu )-1/q))\right.
\notag \\
&&\left. -\frac{qs}{\pi }\mathrm{sgn}(\mu )\sum_{\delta =\pm 1}\delta \cos
(q\pi (1/2-\delta \alpha _{0}))\int_{0}^{\infty }dy\frac{\cosh (\left(
q+2\delta \alpha _{0}q-s\delta \mathrm{sgn}(\mu )\right) y)}{\cosh
(2qy)-\cos (q\pi )}\right] .  \label{FClowT}
\end{eqnarray}%
Here, the term with $l=0$ is the contribution of the Minkowskian part $%
\langle \bar{\psi}\psi \rangle ^{(M)}$. Hence, for $|\mu |<m$, at low
temperatures the thermal contribution in the FC is suppressed by the factor $%
e^{-\beta (m-|\mu |)}$.

The representation (\ref{FC}) is not well adapted for the investigation of
the high temperature limit. An alternative representation, is obtained by
using the relations $(-1)^{n}\cosh (n\beta \mu )=\cos (nb)$ and $%
(-1)^{n}n\sinh (n\beta \mu )=\partial _{\mu }\cos (nb)/\beta $, with $b=\pi
+i\beta \mu $, and then the formula \cite{Bell09}%
\begin{eqnarray}
\sideset{}{'}{\sum}_{n=0}^{\infty }\cos (nb)f_{\nu }(m\sqrt{\beta
^{2}n^{2}+a^{2}}) &=&\frac{(2\pi )^{1/2}}{2\beta m^{2\nu }}\sum_{n=-\infty
}^{+\infty }\left[ (2\pi n+b)^{2}\beta ^{-2}+m^{2}\right] ^{\nu -1/2}  \notag
\\
&&\times f_{\nu -1/2}\left( a\sqrt{(2\pi n+b)^{2}\beta ^{-2}+m^{2}}\right) ,
\label{rep}
\end{eqnarray}%
for the series over $n$ in (\ref{FC}). With these transformations, the FC is
presented as%
\begin{eqnarray}
\langle \bar{\psi}\psi \rangle &=&\langle \bar{\psi}\psi \rangle ^{(M)}-%
\frac{2mT}{\pi }\sum_{n=-\infty }^{\infty }\left\{ \left[
\sum_{l=1}^{p}(-1)^{l}sc_{l}\cos (2\pi l\alpha _{0})\right. \right.  \notag
\\
&&\left. \times K_{0}\left( 2rs_{l}b_{n}\right) -\frac{q}{\pi }%
\int_{0}^{\infty }dy\,\frac{sf_{1}(q,\alpha _{0},y)K_{0}\left( 2rb_{n}\cosh
y\right) }{\cosh (2qy)-\cos (q\pi )}\right]  \notag \\
&&+\frac{\mu -i\pi (2n+1)T}{m}\left[ \sum_{l=1}^{p}(-1)^{l}s_{l}\sin (2\pi
l\alpha _{0})K_{0}\left( 2rs_{l}b_{n}\right) \right.  \notag \\
&&\left. \left. -\frac{q}{\pi }\int_{0}^{\infty }dy\,\frac{f_{2}(q,\alpha
_{0},y)K_{0}\left( 2rb_{n}\cosh y\right) }{\cosh (2qy)-\cos (q\pi )}\right]
\right\} ,  \label{FC13}
\end{eqnarray}%
where we have introduced the notation
\begin{equation}
b_{n}=\{[\pi (2n+1)T+i\mu ]^{2}+m^{2}\}^{1/2}.  \label{bn}
\end{equation}%
In the case of zero chemical potential, $\mu =0$, the expression (\ref{FC13}%
) is simplified to
\begin{eqnarray}
\langle \bar{\psi}\psi \rangle &=&\frac{smT}{\pi }\left\{ \ln (1+e^{-m\beta
})-4\sum_{n=0}^{\infty }\left[ \sum_{l=1}^{p}(-1)^{l}c_{l}\cos (2\pi l\alpha
_{0})\right. \right.  \notag \\
&&\left. \left. \times K_{0}(2rs_{l}b_{n}^{(0)})-\frac{q}{\pi }%
\int_{0}^{\infty }dy\,\frac{f_{1}(q,\alpha _{0},y)K_{0}(2rb_{n}^{(0)}\cosh y)%
}{\cosh (2qy)-\cos (q\pi )}\right] \right\} ,  \label{FC13mu0}
\end{eqnarray}%
with $b_{n}^{(0)}=\sqrt{[\pi (2n+1)T]^{2}+m^{2}}$.

In the high temperature limit, $Tr\gg 1$, the dominant contribution to the
series over $n$ in (\ref{FC13}) comes from the terms with $n=0$ and $n=-1$
and from the part containing the factor $\mu -i\pi (2n+1)T$. In the case $%
q>2 $ the leading contribution corresponds to the term with $l=1$ and we
obtain%
\begin{equation}
\langle \bar{\psi}\psi \rangle \approx \langle \bar{\psi}\psi \rangle
^{(M)}-2\sqrt{s_{1}/r}\sin (2\pi \alpha _{0})\sin (2rs_{1}\mu
)T^{3/2}e^{-2\pi s_{1}Tr}.  \label{FChighT}
\end{equation}%
For $q<2$ the sums over $l$ are absent. At high temperatures the dominant
contribution to the integrals in (\ref{FC13}) comes from the region near the
lower limit of integration. Assuming that $T\gg 1/[\pi r\sin ^{2}(\pi q/2)]$
($q$ is not too close to 2), to the leading order one finds%
\begin{equation}
\langle \bar{\psi}\psi \rangle \approx \langle \bar{\psi}\psi \rangle ^{(M)}-%
\frac{q\sin \left( q\pi \alpha _{0}\right) }{\pi r\sin (\pi q/2)}\sin (2\mu
r)Te^{-2\pi Tr},  \label{FChighT2}
\end{equation}%
and the suppression is stronger. The high-temperature asymptotic for the
Minkowskian part is given by%
\begin{equation}
\langle \bar{\psi}\psi \rangle ^{(M)}\approx \frac{smT}{2\pi }\ln 2,
\label{FChighTM}
\end{equation}%
and, at high temperatures, the FC is dominated by this part. The
contributions induced by the conical defect and the magnetic flux are
exponentially small for points not close to the origin.

For the investigation of the FC at large distances from the origin it is
convenient to use the representation (\ref{FC13}). The corresponding
procedure is similar to that for the high temperature asymptotic and the
dominant contribution comes from the terms with $n=0$ and $n=-1$.
Considering the asymptotic expression for the Macdonald function with large
arguments, for $q>2$ we obtain%
\begin{equation}
\langle \bar{\psi}\psi \rangle \approx \langle \bar{\psi}\psi \rangle
^{(M)}+2\sqrt{\pi s_{1}/r}\sin (2\pi \alpha _{0})T^{2}\mathrm{Im}\left(
b_{0}^{-1/2}e^{-2rs_{1}b_{0}}\right) .  \label{FClarg1}
\end{equation}%
where $b_{0}=\sqrt{(\pi T+i\mu )^{2}+m^{2}}$. In the case $q<2$, assuming
that $\sin (\pi q/2)\gg q/r\sqrt{T^{2}+m^{2}}$, the asymptotic expression is
given by
\begin{equation}
\langle \bar{\psi}\psi \rangle \approx \langle \bar{\psi}\psi \rangle ^{(M)}+%
\frac{qT\sin \left( q\pi \alpha _{0}\right) }{\pi r\sin (\pi q/2)}\mathrm{Re}%
\left[ \left( \mu -i\pi T\right) e^{-2rb_{0}}/b_{0}\right] .  \label{Fclarg2}
\end{equation}%
In both the cases, the parts in the FC induced by the conical defect and by
the magnetic flux are exponentially small and in the leading order we have $%
\langle \bar{\psi}\psi \rangle \approx \langle \bar{\psi}\psi \rangle ^{(M)}$%
.

Now let us turn to the case $|\mu |>m$. For $\mu >m$ ($\mu <-m$) the
contribution of the antiparticles (particles) to the FC is evaluated in a
way similar to that described above and the corresponding expression is
given by (\ref{FCpm5}) with the lower (upper) sign. The total FC is
expressed as%
\begin{eqnarray}
\langle \bar{\psi}\psi \rangle &=&\langle \bar{\psi}\psi \rangle
_{0}+\langle \bar{\psi}\psi \rangle _{\pm }+\frac{1}{\pi }\int_{0}^{\infty
}d\gamma \frac{\gamma }{e^{\beta (E-|\mu |)}+1}  \notag \\
&&\times \left\{ \sideset{}{'}{\sum}_{l=0}^{p}(-1)^{l}\left[ \frac{sm}{E}%
c_{l}\cos (2\pi l\alpha _{0})\mp s_{l}\sin (2\pi l\alpha _{0})\right]
J_{0}(2\gamma rs_{l})\right.  \notag \\
&&\left. -\frac{q}{\pi }\int_{0}^{\infty }dy\frac{\frac{sm}{E}f_{1}(q,\alpha
_{0},y)\mp f_{2}(q,\alpha _{0},y)}{\cosh (2qy)-\cos (q\pi )}J_{0}(2\gamma
r\cosh y)\right\} .  \label{FCv2}
\end{eqnarray}%
where the upper and lower signs correspond to $\mu <-m$ and $\mu >m$,
respectively. Introducing the Fermi momentum $p_{0}=\sqrt{\mu ^{2}-m^{2}}$,
the integration over $\gamma $ in the last term can be divided into two
regions with $\gamma <p_{0}$ and $\gamma >p_{0}$. The expansion (\ref%
{Expansion}) can be further applied to the integral over the second region.
At high temperatures, $T\gg |\mu |$, the dominant contribution to the FC
comes from the states with the energies $E\gg |\mu |$ and the asymptotic
estimates considered above for the case $|\mu |<m$ are still valid.

Compared to the case $|\mu |<m$, the situation in the range $|\mu |>m$ is
completely different at low temperatures. In the limit $T\rightarrow 0$, in
the part coming from the particles or antiparticles, the contribution of the
states with $E\leqslant |\mu |$ survives only and we find%
\begin{eqnarray}
\langle \bar{\psi}\psi \rangle _{T=0} &=&\langle \bar{\psi}\psi \rangle _{0}+%
\frac{p_{0}^{2}}{\pi }\left\{ \sideset{}{'}{\sum}_{l=0}^{p}(-1)^{l}\right.
\notag \\
&&\times \left[ sc_{l}\cos (2\pi l\alpha _{0})g_{1}(p_{0}rs_{l})+\mathrm{sgn}%
(\mu )s_{l}\sin (2\pi l\alpha _{0})g_{2}(p_{0}rs_{l})\right]  \notag \\
&&\left. -\frac{q}{\pi }\int_{0}^{\infty }dy\frac{\sum_{l=1}^{2}s^{l}(%
\mathrm{sgn}(\mu ))^{l-1}f_{l}(q,\alpha _{0},y)g_{l}(p_{0}r\cosh y)}{\cosh
(2qy)-\cos (q\pi )}\right\} .  \label{FCT0}
\end{eqnarray}%
with the notations%
\begin{equation}
g_{1}(u)=\frac{m}{p_{0}}\int_{0}^{1}dx\frac{xJ_{0}(2ux)}{\sqrt{%
x^{2}+m^{2}/p_{0}^{2}}},\;g_{2}(u)=\frac{J_{1}(2u)}{2u}.  \label{g12}
\end{equation}%
The second term in the right-hand side of (\ref{FCT0}) is the contribution
of particles (for $\mu >m$) or antiparticles (for $\mu <-m$) filling the
states with the energies $m\leqslant E\leqslant |\mu |$. This term is finite
on the apex for $2|\alpha _{0}|<1-1/q$ and diverges as $1/r^{1-2\rho }$ for $%
2|\alpha _{0}|>1-1/q$. The vacuum contribution near the apex behaves as $1/r$
and it dominates. At large distances from the apex, $p_{0}r\gg 1$, the
corrections induced by the conical defect and by the magnetic flux are
dominated by the part coming from the particles or antiparticles (the second
term in the right-hand side of (\ref{FCT0}) with the contribution of $l=0$
excluded). This part decays oscillatory with the amplitude decreasing as $%
1/r^{3/2}$. Note that the decay of the vacuum part for a massive field is
exponential. In the case of a massless field, $m=0$, the vacuum part of the
FC vanishes and the formula (\ref{FCT0}) simplifies to
\begin{eqnarray}
\langle \bar{\psi}\psi \rangle _{T=0} &=&\mathrm{sgn}(\mu )\frac{\mu ^{2}}{%
\pi }\left[ \sideset{}{'}{\sum}_{l=0}^{p}(-1)^{l}s_{l}\sin (2\pi l\alpha
_{0})g_{2}(|\mu |rs_{l})\right.  \notag \\
&&\left. -\frac{q}{\pi }\int_{0}^{\infty }dy\frac{f_{2}(q,\alpha
_{0},y)g_{2}(|\mu |r\cosh y)}{\cosh (2qy)-\cos (q\pi )}\right] .
\label{FCT0m0}
\end{eqnarray}%
In this case, the only nonzero contribution comes from particles or
antiparticles and the FC\ is an odd function of both the chemical potential
and the magnetic flux.

In the discussion above we have assumed that the field $\psi (x)$ is
periodic along the azimuthal direction. We may consider a more general case
where the spinor field obeys the quasiperiodicity condition
\begin{equation}
\psi (t,r,\phi +\phi _{0})=e^{i\chi }\psi (t,r,\phi ),  \label{periodcond}
\end{equation}%
with a constant phase $\chi $. The corresponding expressions for the
expectation values are obtained from those presented above (and in what
follows) with $\alpha $ given by the formula%
\begin{equation}
\alpha =\chi /(2\pi )+eA/q.  \label{Replalf}
\end{equation}%
Note that $\chi $ and $A$ are changed by a gauge transformation whereas
their combination in the right-hand side of (\ref{Replalf}) is gauge
invariant.

\section{Charge density}

\label{sec:Charge}

In this and in the following section we shall investigate the expectation
value of the fermionic current density given by $\langle j^{\nu }\rangle =e\
\mathrm{tr\,}[\hat{\rho}\bar{\psi}(x)\gamma ^{\nu }\psi (x)]$. Similarly to
the case of the FC, the current density is decomposed as
\begin{equation}
\langle j^{\nu }\rangle =\langle j^{\nu }\rangle _{0}+\langle j^{\nu
}\rangle _{+}+\langle j^{\nu }\rangle _{-},  \label{jnudec}
\end{equation}%
where%
\begin{equation}
\langle j^{\nu }\rangle _{0}=e\sum_{\sigma }\bar{\psi}_{\sigma
}^{(-)}(x)\gamma ^{\nu }\psi _{\sigma }^{(-)}(x),  \label{current3}
\end{equation}%
is the vacuum expectation value and%
\begin{equation}
\langle j^{\nu }\rangle _{\pm }=\pm e\sum_{\sigma }\frac{\bar{\psi}_{\sigma
}^{(\pm )}\gamma ^{\nu }\psi _{\sigma }^{(\pm )}}{e^{\beta (E_{\sigma }\mp
\mu )}+1}.  \label{current4}
\end{equation}%
The terms $\langle j^{\nu }\rangle _{+}$ and $\langle j^{\nu }\rangle _{-}$
are the contributions to the current density coming from particles and
antiparticles. The vacuum expectation value has been investigated in \cite%
{Beze10} and here we will be focused with the finite temperature effects.
The details of the calculations are similar to those for the FC and the main
steps only will be given.

We start with the charge density that corresponds to the component $\nu =0$
in (\ref{jnudec}). By using the mode functions (\ref{Modes}), for the
contributions from the particles and antiparticles we get the representation%
\begin{equation}
\langle j^{0}\rangle _{\pm }=\pm \frac{eq}{4\pi }\sum_{j}\int_{0}^{\infty
}d\gamma \frac{\gamma }{e^{\beta (E\mp \mu )}+1}\left\{ J_{\beta
_{j}}^{2}(\gamma r)+J_{\beta _{j}+\epsilon _{\beta _{j}}}^{2}(\gamma r)\pm
\frac{sm}{E}\left[ J_{\beta _{j}}^{2}(\gamma r)-J_{\beta _{j}+\epsilon
_{\beta _{j}}}^{2}(\gamma r)\right] \right\} .  \label{j0pm}
\end{equation}%
As we could expect $\pm \langle j^{0}\rangle _{\pm }/e>0$. By the
transformations similar to those for the FC, one finds the following
expression%
\begin{eqnarray}
\langle j^{0}\rangle _{\pm } &=&\langle j^{0}\rangle _{\pm }^{(M)}+\frac{e}{%
\pi }\int_{0}^{\infty }d\gamma \,\frac{\gamma }{e^{\beta (E\mp \mu )}+1}
\notag \\
&&\left\{ \sum_{l=1}^{p}(-1)^{l}\left[ \frac{sm}{E}s_{l}\sin (2\pi l\alpha
_{0})\pm c_{l}\cos (2\pi l\alpha _{0})\right] J_{0}(2\gamma rs_{l})\right.
\notag \\
&&\left. -\frac{q}{\pi }\int_{0}^{\infty }dy\,\frac{\frac{sm}{E}%
f_{2}(q,\alpha _{0},2y)\pm f_{1}(q,\alpha _{0},2y)}{\cosh (2qy)-\cos (q\pi )}%
J_{0}(2\gamma r\cosh y)\right\} ,  \label{j0pm12}
\end{eqnarray}%
with the Minkowskian part%
\begin{equation}
\langle j^{0}\rangle _{\pm }^{(M)}=\pm \frac{e}{2\pi }\int_{0}^{\infty
}d\gamma \,\frac{\gamma }{e^{\beta (E\mp \mu )}+1}.  \label{j0pmM1}
\end{equation}%
For a massless field and in the case of the zero chemical potential, $\mu =0$%
, the contributions from the particles and antiparticles to the total charge
density cancel each other: $\langle j^{0}\rangle _{+}=-\langle j^{0}\rangle
_{-}$. This is not the case for a massive field.

In the case $|\mu |\leqslant m$, by using the relation (\ref{Expansion}), we
get
\begin{eqnarray}
\langle j^{0}\rangle _{\pm } &=&\langle j^{0}\rangle _{\pm }^{(M)}-\frac{%
\sqrt{2}em^{2}}{\pi ^{3/2}}\sum_{n=1}^{\infty }(-1)^{n}e^{\pm n\beta \mu }
\notag \\
&&\times \left\{ \sum_{l=1}^{p}(-1)^{l}\left[ ss_{l}\sin (2\pi l\alpha
_{0})f_{1/2}\left( c_{nl}\right) \pm mn\beta c_{l}\cos (2\pi l\alpha
_{0})f_{3/2}\left( c_{nl}\right) \right] \right.  \notag \\
&&\left. -\frac{q}{\pi }\int_{0}^{\infty }dy\left[ \frac{sf_{2}(q,\alpha
_{0},y)f_{1/2}\left( c_{n}(y)\right) }{\cosh (2qy)-\cos (q\pi )}\pm \frac{%
mn\beta f_{1}(q,\alpha _{0},y)f_{3/2}\left( c_{n}(y)\right) }{\cosh
(2qy)-\cos (q\pi )}\right] \right\} ,  \label{j0pm1}
\end{eqnarray}%
with
\begin{equation}
\langle j^{0}\rangle _{\pm }^{(M)}=\mp \frac{em^{3}\beta }{\sqrt{2}\pi ^{3/2}%
}\sum_{n=1}^{\infty }(-1)^{n}ne^{\pm n\beta \mu }f_{3/2}\left( mn\beta
\right) .  \label{j0pmM}
\end{equation}%
The vacuum expectation value of the charge density is given by the
expression \cite{Beze10}%
\begin{eqnarray}
\langle j^{0}\rangle _{0} &=&-\frac{sem}{2\pi r}\left[
\sum_{l=1}^{p}(-1)^{l}\sin (2\pi l\alpha _{0})e^{-2mrs_{l}}\right.  \notag \\
&&\left. -\frac{q}{\pi }\int_{0}^{\infty }dy\,\frac{f_{2}(q,\alpha _{0},y)}{%
\cosh (2qy)-\cos (q\pi )}\frac{e^{-2mr\cosh y}}{\cosh y}\right] .
\label{j00}
\end{eqnarray}%
For the case of a massless field, because of the condition $|\mu |\leqslant
m $, in (\ref{j0pm1}) we should also assume $\mu =0$. By taking into account
that $f_{\nu }(x)\approx 2^{\nu -1}\Gamma (\nu )x^{-2\nu }$ for$%
\;x\rightarrow 0$, we find the expression%
\begin{eqnarray}
\langle j^{0}\rangle _{\pm } &=&\pm \frac{\pi eT^{2}}{24}\mp \frac{e}{\pi T}%
\sum_{n=1}^{\infty }(-1)^{n}n\left[ \sum_{l=1}^{p}\frac{(-1)^{l}c_{l}\cos
(2\pi l\alpha _{0})}{(n^{2}\beta ^{2}+4r^{2}s_{l}^{2})^{3/2}}\right.  \notag
\\
&&\left. -\frac{q}{\pi }\int_{0}^{\infty }dy\,\frac{f_{1}(q,\alpha
_{0},y)(n^{2}\beta ^{2}+4r^{2}\cosh ^{2}y)^{-3/2}}{\cosh (2qy)-\cos (q\pi )}%
\right] ,  \label{j0pmm0}
\end{eqnarray}%
where the first term in the right-hand side is the Minkowskian part. In this
case the vacuum charge density vanishes. Note that for a massless field the
charge densities (\ref{j0pmm0}) do not depend on the parameter $s$. They are
even functions of $\alpha _{0}$.

Summing the contributions from the vacuum expectation value and from
particles and antiparticles, for the total charge density, in the case $|\mu
|\leqslant m$, we obtain%
\begin{eqnarray}
\langle j^{0}\rangle &=&\langle j^{0}\rangle ^{(M)}-\frac{2^{3/2}em^{2}}{\pi
^{3/2}}\left\{ \sideset{}{'}{\sum}_{n=0}^{\infty }(-1)^{n}s\cosh {(n\beta
\mu )}\left[ \sum_{l=1}^{p}(-1)^{l}\right. \right.  \notag \\
&&\times \left. s_{l}\sin (2\pi l\alpha _{0})f_{1/2}\left( c_{nl}\right) -%
\frac{q}{\pi }\int_{0}^{\infty }dy\,\frac{f_{2}(q,\alpha
_{0},y)f_{1/2}\left( c_{n}(y)\right) }{\cosh (2qy)-\cos (q\pi )}\right]
\notag \\
&&+m\beta \sum_{n=1}^{\infty }(-1)^{n}n\sinh {(n\beta \mu )}\left[
\sum_{l=1}^{p}(-1)^{l}c_{l}\cos (2\pi l\alpha _{0})f_{3/2}\left(
c_{nl}\right) \right.  \notag \\
&&\left. \left. -\frac{q}{\pi }\int_{0}^{\infty }dy\,\frac{f_{1}(q,\alpha
_{0},y)f_{3/2}\left( c_{n}(y)\right) }{\cosh (2qy)-\cos (q\pi )}\right]
\right\} ,  \label{j0}
\end{eqnarray}%
with the Minkowskian term%
\begin{equation}
\langle j^{0}\rangle ^{(M)}=-\frac{\sqrt{2}em^{3}}{\pi ^{3/2}T}%
\sum_{n=1}^{\infty }(-1)^{n}n\sinh {(n\beta \mu )}f_{3/2}\left( mn\beta
\right) .  \label{j0M}
\end{equation}%
Note that, in the case of zero chemical potential, the Minkowskian part in
the charge density vanishes as a consequence of the cancellation of the
contributions coming from particles and antiparticles. The magnetic flux
acts on these contributions in different ways and there is no such a
cancellation for the topological part. The $n=0$ term in (\ref{j0})
corresponds to the vacuum expectation value of the charge density. In the
case of a massless field with the zero chemical potential the charge density
vanishes. Note that, for a massive field, the charge density has indefinite
parity with respect to the change of the sign for $\alpha _{0}$, $\alpha
_{0}\rightarrow -\alpha _{0}$, and for $\mu $, $\mu \rightarrow -\mu $. This
is related to the fact that, in the presence of the mass term, the model
under consideration is not invariant under the $T$- and $P$-transformations.

\begin{figure}[tbph]
\begin{center}
\epsfig{figure=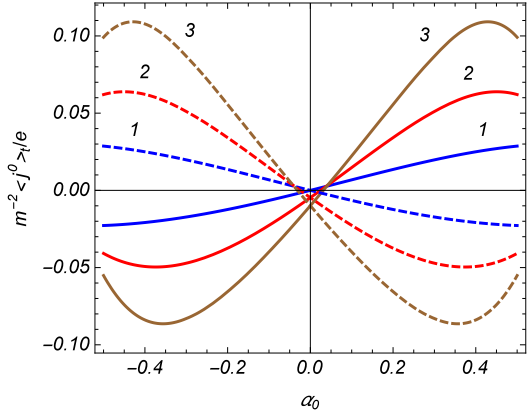,width=7.cm,height=5.5cm}
\end{center}
\caption{Topological part in the charge density as a function of the
parameter $\protect\alpha _{0}$ for fields realizing the irreducible
representations with $s=1$ (full curves) and $s=-1$ (dashed curves). The
numbers near the curves are the corresponding values of $q$. The graphs are
plotted for $\protect\mu /m=0.25$, $mr=0.5$, and $T/m=0.5$.}
\label{fig1}
\end{figure}

The quantity $\langle j^{0}\rangle _{\mathrm{t}}=\langle j^{0}\rangle
-\langle j^{0}\rangle ^{(M)}$ gives the contribution to the charge density
coming from the planar angle deficit and from the magnetic flux. For brevity
we shall call it as topological part. In Figure \ref{fig1} we have plotted
this part versus the parameter $\alpha _{0}$. Recall that the charge density
is a periodic function of $\alpha _{0}$ with the period 1. The full and
dashed curves correspond to the irreducible representation with $s=1$ and $%
s=-1$, respectively. The numbers near the curves are the values of $q$. The
graphs are plotted for $\mu /m=0.25$, $mr=0.5$, $T/m=0.5$. For this example,
the contribution of the terms in (\ref{j0}) containing the factor $s$
dominates. These terms are odd with respect to the reflection $\alpha
_{0}\rightarrow -\alpha _{0}$. Note that for the same values of $\mu /m=0.25$%
, $T/m=0.5$, for the Minkowskian part one has $\langle j^{0}\rangle
^{(M)}\approx 0.015em^{2}$.

The general formula (\ref{j0}) is simplified in special cases. For integer
values of $q$ and for the values of $\alpha _{0}$ given by (\ref{alfsp}),
the integral terms vanish. In the absence of the magnetic flux and for
general values of $q$, the charge density is given by%
\begin{eqnarray}
\langle j^{0}\rangle &=&-\frac{em^{3}\beta }{(\pi /2)^{3/2}}%
\sum_{n=1}^{\infty }(-1)^{n}n\sinh {(n\beta \mu )}\left[ \sideset{}{'}{\sum}%
_{l=0}^{p}(-1)^{l}c_{l}f_{3/2}\left( c_{nl}\right) \right.  \notag \\
&&\left. +\frac{2q}{\pi }\cos (q\pi /2)\int_{0}^{\infty }dy\frac{\sinh
y\sinh (qy)f_{3/2}\left( c_{n}(y)\right) }{\cosh (2qy)-\cos (q\pi )}\right] ,
\label{j0flux0}
\end{eqnarray}%
and it is an odd function of the chemical potential. For the charge density
in the background of Minkowski spacetime ($q=1$) in the presence of a
magnetic flux we get%
\begin{eqnarray}
\langle j^{0}\rangle &=&\langle j^{0}\rangle ^{(M)}+\frac{\sqrt{2}em^{2}}{%
\pi ^{5/2}}\sin (\pi \alpha _{0})\sideset{}{'}{\sum}_{n=0}^{\infty }(-1)^{n}
\notag \\
&&\times \left[ s\cosh {(n\beta \mu )}\int_{0}^{\infty }dy\cosh (2\alpha
_{0}y)f_{1/2}\left( c_{n}(y)\right) \right.  \notag \\
&&\left. -nm\beta \sinh {(n\beta \mu )}\int_{0}^{\infty }dy\tanh y\sinh
(2\alpha _{0}y)f_{3/2}\left( c_{n}(y)\right) \right] .  \label{j0q1}
\end{eqnarray}

For $2|\alpha _{0}|<1-1/q$, the finite temperature part in the charge
density is finite at the cone apex. This part is given by the right-hand
side of (\ref{j0}), excluding the term $n=0$. The corresponding expression
is obtained by the direct substitution $r=0$. In the case $2|\alpha
_{0}|>1-1/q$, the analysis similar to that for the FC leads to the following
asymptotic expression%
\begin{eqnarray}
\langle j^{0}\rangle &\approx &\langle j^{0}\rangle _{0}+\frac{em^{2}q\Gamma
(1/2-\rho )}{2^{\rho }\pi ^{5/2}(mr)^{1-2\rho }}\cos \left( \pi \rho \right)
\sum_{n=1}^{\infty }(-1)^{n}  \notag \\
&&\times \left[ \mathrm{sgn}(\alpha _{0})s\cosh {(n\beta \mu )}f_{\rho
}(mn\beta )-m\beta n\sinh {(n\beta \mu )}f_{\rho +1}(mn\beta )\right] ,
\label{j0r0}
\end{eqnarray}%
and the charge density diverges as $1/r^{1-2\rho }$. The vacuum expectation
value, $\langle j^{0}\rangle _{0}$, diverges in the limit $r\rightarrow 0$
as $1/r$ and near the apex it dominates in the total charge density. Though
the charge density diverges on the apex, this divergence is integrable and
the total charge induced by the planar angle deficit and by the magnetic
flux is finite (see below).

Now, let us analyze the limits of low and high temperatures. At low
temperatures, for a fixed value of $mr$, the main contribution to the finite
temperature part of the charge density comes from the $n=1$ term in (\ref{j0}%
) and it is given by%
\begin{eqnarray}
\langle j^{0}\rangle &\approx &\langle j^{0}\rangle _{0}+\frac{em}{\pi \beta
}\frac{\mathrm{sgn}(\mu )}{e^{\beta (m-{|\mu |})}}\left[ \sideset{}{'}{\sum}%
_{l=0}^{p}(-1)^{l}\right.  \notag \\
&&\times \left. \cos (\pi l(1/q-2\mathrm{sgn}(\mu )s\alpha _{0}))-\frac{q}{%
\pi }\int_{0}^{\infty }dy\frac{f(q,\mathrm{sgn}(\mu )s\alpha _{0},2y)}{\cosh
(2qy)-\cos (q\pi )}\right] ,  \label{j0smT}
\end{eqnarray}%
where the $l=0$ term corresponds to the Minkowskian part. In this limit one
has an exponential suppression of the thermal effects.

To evaluate the charge density at high temperatures, it is convenient to use
another representation that is obtained from (\ref{j0}) in the way similar
to that we have used for (\ref{FC13}). The new representation has the form
\begin{eqnarray}
\langle j^{0}\rangle &=&\langle j^{0}\rangle ^{(M)}-\frac{2emT}{\pi }%
\sum_{n=-\infty }^{\infty }\left\{ s\sum_{l=1}^{p}(-1)^{l}s_{l}\sin (2\pi
l\alpha _{0})K_{0}(2rs_{l}b_{n})\right.  \notag \\
&&-\frac{sq}{\pi }\int_{0}^{\infty }dy\,\frac{f_{2}(q,\alpha
_{0},y)K_{0}(2rb_{n}\cosh y)}{\cosh (2qy)-\cos (q\pi )}  \notag \\
&&+\frac{\mu -i\pi (2n+1)T}{m}\left[ \sum_{l=1}^{p}(-1)^{l}c_{l}\cos (2\pi
l\alpha _{0})K_{0}(2rs_{l}b_{n})\right.  \notag \\
&&\left. \left. -\frac{q}{\pi }\int_{0}^{\infty }dy\,\frac{f_{1}(q,\alpha
_{0},y)K_{0}(2rb_{n}\cosh y)}{\cosh (2qy)-\cos (q\pi )}\right] \right\} ,
\label{j02}
\end{eqnarray}%
where $b_{n}$ is given by (\ref{bn}). For a field with zero chemical
potential we get
\begin{eqnarray}
\langle j^{0}\rangle &=&-\frac{4semT}{\pi }\sum_{n=0}^{\infty }\left[
\sum_{l=1}^{p}(-1)^{l}s_{l}\sin (2\pi l\alpha
_{0})K_{0}(2rs_{l}b_{n}^{(0)})\right.  \notag \\
&&\left. -\frac{q}{\pi }\int_{0}^{\infty }dy\,\frac{f_{2}(q,\alpha
_{0},y)K_{0}(2rb_{n}^{(0)}\cosh y)}{\cosh (2qy)-\cos (q\pi )}\right] ,
\label{j02mu0}
\end{eqnarray}%
with the same $b_{n}^{(0)}$ as in (\ref{FC13mu0}).

In the limit of high temperatures, the dominant contributions comes from the
terms with $n=0$ and $n=-1$. For the case $q>2$ the leading term corresponds
to the $l=1$ term and one finds%
\begin{equation}
\langle j^{0}\rangle =\langle j^{0}\rangle ^{(M)}-\frac{2ec_{1}}{\sqrt{rs_{1}%
}}\cos (2\pi \alpha _{0})\sin (2rs_{1}\mu )T^{3/2}e^{-2\pi rs_{1}T}.
\label{j0highT}
\end{equation}%
For $q<2$ the summs over $l$ in (\ref{j02}) are absent and the dominant
contribution to the integrals over $y$ comes from the region near the lower
limit of integration. In this case the effects induced by the conical defect
and magnetic flux are suppressed by the factor $e^{-2\pi rT}$. In all cases,
for $\mu \neq 0$, at high temperatures the charge density is dominated by
the Minkowskian part $\langle j^{0}\rangle ^{(M)}$ with the asymptotic%
\begin{equation}
\langle j^{0}\rangle ^{(M)}\approx \frac{e\mu T}{\pi }\ln 2.
\label{j0MhighT}
\end{equation}%
In the case of zero chemical potential the Minkowskian part vanishes. For
points not too close to the cone apex, the contributions induced by the
conical defect and by the magnetic flux are exponentially small.

\begin{figure}[tbph]
\begin{center}
\epsfig{figure=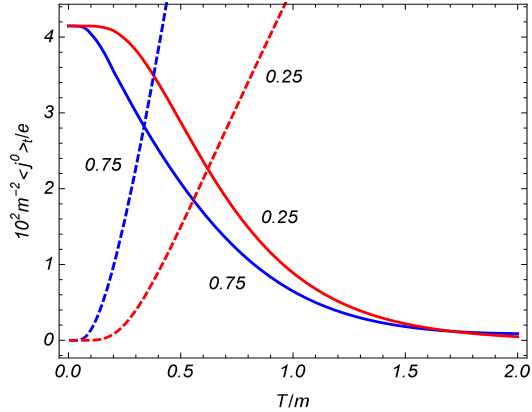,width=7.cm,height=5.5cm}
\end{center}
\caption{Topological part in the charge density versus the temperature for
the irreducible representation with $s=1$ (full curves). The graphs are
plotted for $q=1.5$, $\protect\alpha _{0}=0.25$, $mr=0.5$. The dashed curves
present the charge density in the Minkowski spacetime. The numbers near the
curves are the values of $\protect\mu /m$.}
\label{fig2}
\end{figure}

The full curves in Figure \ref{fig2} present the dependence of the
topological part in the charge density on the temperature. The dashed curves
correspond to the charge density in Minkowski spacetime in the absence of
the magnetic flux, $m^{-2}\langle j^{0}\rangle ^{(M)}/e$. The numbers near
the curves are the values of the ratio $\mu /m$. The graphs are plotted for $%
q=1.5$, $\alpha _{0}=0.25$, $mr=0.5$, and for the irreducible representation
with $s=1$.

In the same way as for the FC, the representation (\ref{j02}) also is
convenient for the investigation of the charge density at large distances
from the origin. As before, the corresponding procedure is similar to that
for the high temperature asymptotic. The main contributions come from the
terms with $n=0$ and $n=-1$. Considering $q>2$, one obtains%
\begin{equation}
\langle j^{0}\rangle \approx \langle j^{0}\rangle ^{(M)}+\frac{2eT}{\sqrt{%
\pi rs_{1}}}\mathrm{Re}\left[ \frac{sms_{1}\sin (2\pi \alpha _{0})+(\mu
-i\pi T)c_{1}\cos (2\pi \alpha _{0})}{b_{0}^{1/2}e^{2rs_{1}b_{0}}}\right] .
\label{j0larger}
\end{equation}%
For $q<2$, the corresponding asymptotic expression is given by%
\begin{equation}
\langle j^{0}\rangle \approx \langle j^{0}\rangle ^{(M)}+\frac{esqmT}{\pi r}%
\frac{\sin (q\pi \alpha _{0})}{\sin (q\pi /2)}\mathrm{Re}\left( \frac{%
e^{-2rb_{0}}}{b_{0}}\right) .  \label{j0largr2}
\end{equation}%
In both cases, the parts induced by the conical defect and the magnetic flux
are exponentially small and to the leading order we have $\langle
j^{0}\rangle \approx \langle j^{0}\rangle ^{(M)}$.

For $\mu <-m$ ($\mu >m$) the contribution of the particles (antiparticles)
to the charge density is still given by (\ref{j0pm1}) with the upper (lower)
sign. The total charge density is expressed as%
\begin{eqnarray}
\langle j^{0}\rangle &=&\langle j^{0}\rangle _{0}+\langle j^{0}\rangle _{\pm
}+\frac{e}{\pi }\int_{0}^{\infty }d\gamma \,\frac{\gamma }{e^{\beta (E-|\mu
|)}+1}  \notag \\
&&\left\{ \sideset{}{'}{\sum}_{l=0}^{p}(-1)^{l}\left[ \frac{sm}{E}s_{l}\sin
(2\pi l\alpha _{0})\mp c_{l}\cos (2\pi l\alpha _{0})\right] J_{0}(2\gamma
rs_{l})\right.  \notag \\
&&\left. -\frac{q}{\pi }\int_{0}^{\infty }dy\,\frac{\frac{sm}{E}%
f_{2}(q,\alpha _{0},2y)\mp f_{1}(q,\alpha _{0},2y)}{\cosh (2qy)-\cos (q\pi )}%
J_{0}(2\gamma r\cosh y)\right\} ,  \label{j0v2}
\end{eqnarray}%
where upper and lower signs correspond to $\mu <-m$ and $\mu >m$,
respectively, and for $\langle j^{0}\rangle _{\pm }$ one has the expression (%
\ref{j0pm1}). Considering the separate regions $\gamma <p_{0}$ and $\gamma
>p_{0}$, in the integral corresponding to the second region we can again use
the expansion (\ref{Expansion}). The leading terms in the high temperature
asymptotic remain the same as for the case $|\mu |\leqslant m$.

At zero temperature and for the case $|\mu |>m$, the only nonzero
contribution to the $\gamma $-integral in (\ref{j0v2}) comes from the region
$\gamma <p_{0}$ and one obtains%
\begin{eqnarray}
\langle j^{0}\rangle _{T=0} &=&\langle j^{0}\rangle _{0}+\frac{ep_{0}^{2}}{%
\pi }\left\{ \sideset{}{'}{\sum}_{l=0}^{p}(-1)^{l}\right.  \notag \\
&&\times \left[ ss_{l}\sin (2\pi l\alpha _{0})g_{1}(p_{0}rs_{l})+\mathrm{sgn}%
(\mu )c_{l}\cos (2\pi l\alpha _{0})g_{2}(p_{0}rs_{l})\right]  \notag \\
&&\left. -\frac{q}{\pi }\int_{0}^{\infty }dy\frac{\sum_{l=1}^{2}s^{l-1}(%
\mathrm{sgn}(\mu ))^{l}f_{l}(q,\alpha _{0},y)g_{3-l}(p_{0}r\cosh y)}{\cosh
(2qy)-\cos (q\pi )}\right\} ,  \label{j0T0}
\end{eqnarray}%
with the functions $g_{1}(u)$ and $g_{2}(u)$ from (\ref{g12}). The
appearance of the second term is related to the presence of antiparticles
(for $\mu <-m$) or particles (for $\mu >m$) in the states having the energy $%
m\leqslant E\leqslant |\mu |$. In the absence of the planar angle deficit
and magnetic flux ($q=1$, $\alpha =0$), the $l=0$ term remains only and for
the charge density we obtain the expression: $\langle j^{0}\rangle
_{T=0}^{(M)}=\mathrm{sgn}(\mu )ep_{0}^{2}/(2\pi )$. Subtracting from the
right-hand side of (\ref{j0T0}) the term $l=0$, we obtain the charge density
induced by the planar angle deficit and by the magnetic flux. At large
distances from the cone apex the decay of this part in the charge density is
oscillatory with the amplitude decreasing as $1/r^{3/2}$. Similar to the
case $|\mu |\leqslant m$, the charge density is finite at the apex for $%
2|\alpha _{0}|<1-1/q$ and diverges in the case $2|\alpha _{0}|>1-1/q$. The
divergence in the second case is integrable, as $1/r^{1-2\rho }$. For a
massless field the vacuum charge density is zero and from (\ref{j0T0m0}) we
get the following expression%
\begin{eqnarray}
\langle j^{0}\rangle _{T=0} &=&\mathrm{sgn}(\mu )\frac{e\mu ^{2}}{\pi }\left[
\sideset{}{'}{\sum}_{l=0}^{p}(-1)^{l}c_{l}\cos (2\pi l\alpha _{0})\right.
\notag \\
&&\times \left. g_{2}(|\mu |rs_{l})-\frac{q}{\pi }\int_{0}^{\infty }dy\frac{%
f_{1}(q,\alpha _{0},y)g_{2}(|\mu |r\cosh y)}{\cosh (2qy)-\cos (q\pi )}\right]
.  \label{j0T0m0}
\end{eqnarray}%
In this case the charge density at zero temperature is the same for both the
irreducible representations of the Clifford algebra. It is an odd function
of the chemical potential and an even function of the magnetic flux.

\begin{figure}[tbph]
\begin{center}
\begin{tabular}{cc}
\epsfig{figure=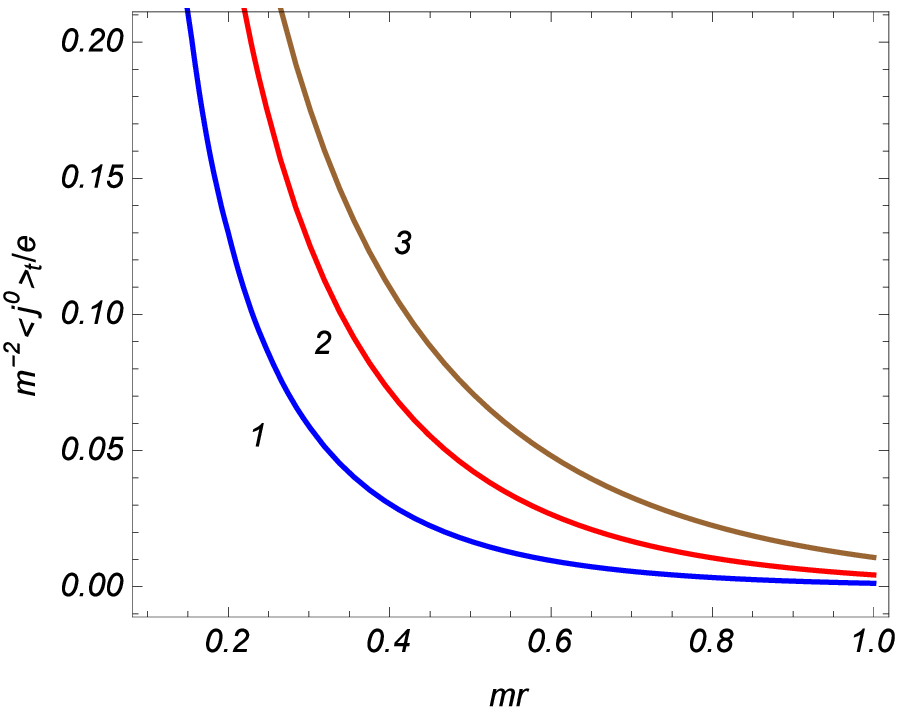,width=7.cm,height=5.5cm} & \quad %
\epsfig{figure=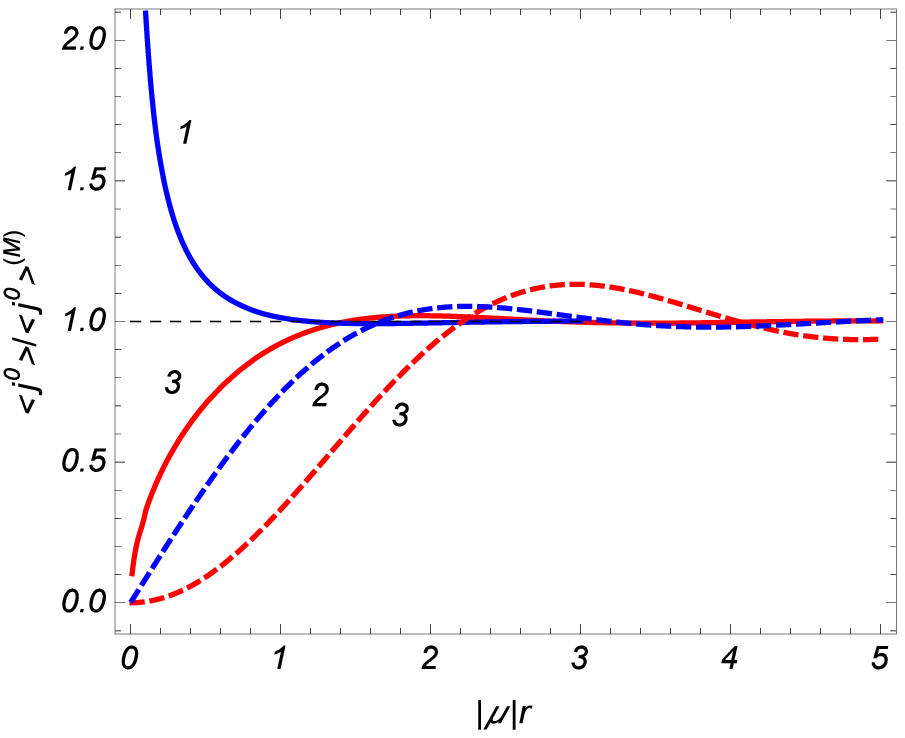,width=7.cm,height=5.5cm}%
\end{tabular}%
\end{center}
\caption{Charge density as a function of the distance from the apex for
different values of the parameter $q$ (the numbers near the graphs). On the
left panel, the topological part is presented for the representation $s=1$
and for $\protect\mu /m=0.25$, $\protect\alpha _{0}=0.25$, $T/m=0.5$. The
right panel gives the dependence of the ratio $\langle j^{0}\rangle /\langle
j^{0}\rangle ^{(M)}$ at zero temperature. The full and dashed curves on this
panel correspond to $\protect\alpha _{0}=0.25$ and $\protect\alpha _{0}=0$.}
\label{fig3}
\end{figure}

In Figure \ref{fig3} we displayed the charge density as a function of the
distance from the cone apex for different values of $q$ (the numbers near
the curves). The left panel presents the topological part in the charge
density for the field realizing the irreducible representation with $s=1$
and for $\mu /m=0.25$, $\alpha _{0}=0.25$, $T/m=0.5$. In the right panel the
ratio $\langle j^{0}\rangle /\langle j^{0}\rangle ^{(M)}$ is plotted at zero
temperature for a massless fermionic field. The full and dashed curves
correspond to $\alpha _{0}=0.25$ and $\alpha _{0}=0$, respectively. Note
that for $\alpha _{0}=0.25$ and $q=2$ one has $\langle j^{0}\rangle
_{T=0}=\langle j^{0}\rangle _{T=0}^{(M)}$.

Let us denote by $\Delta Q$ the total charge induced by the planar angle
deficit and by the magnetic flux:
\begin{equation}
\Delta Q=\int_{0}^{\infty }dr\,r\int_{0}^{\phi _{0}}d\phi \,\left[ \langle
j^{0}\rangle -\langle j^{0}\rangle ^{(M)}\right] .  \label{DelQ}
\end{equation}%
In the case $|\mu |\leqslant m$ we use the representation (\ref{j0}). The
integration over the radial coordinate is done with the help of the formula
\begin{equation}
\int_{0}^{\infty }dr\,rf_{\nu }(\sqrt{a^{2}+b^{2}r^{2}})=f_{\nu -1}(a)/b^{2}.
\label{Int3}
\end{equation}%
After the summation over $n$ we get%
\begin{eqnarray}
\Delta Q &=&\Delta Q_{0}+\frac{e}{2}\sum_{\delta =\pm 1}\frac{\delta }{%
e^{\beta \left( m-\delta \mu \right) }+1}\left[ \frac{1}{q}\sum_{l=1}^{p}%
\frac{(-1)^{l}}{s_{l}^{2}}\right.  \notag \\
&&\times \left. \cos (\pi l\left( 1/q-2s\delta \alpha _{0}\right) )-\frac{1}{%
\pi }\int_{0}^{\infty }dy\,\frac{f(q,s\delta \alpha _{0},2y)\cosh ^{-2}y}{%
\cosh (2qy)-\cos (q\pi )}\right] ,  \label{DelQ1}
\end{eqnarray}%
where%
\begin{equation}
\Delta Q_{0}=-\frac{se}{2q}\left[ \sum_{l=1}^{p}\frac{(-1)^{l}}{s_{l}}\sin
(2\pi l\alpha _{0})-\frac{q}{\pi }\int_{0}^{\infty }dy\,\frac{f_{2}(q,\alpha
_{0},y)\cosh ^{-2}y}{\cosh (2qy)-\cos (q\pi )}\right] ,  \label{DelQ0}
\end{equation}%
is the vacuum charge and the second term in the right-hand side presents the
contribution from particles and antiparticles. These expressions are further
simplified by using the relations (\ref{Rel4}) and (\ref{Rel4b}):
\begin{equation}
\Delta Q=\Delta Q_{0}+\frac{e}{2q}\sum_{\delta =\pm 1}\frac{\delta }{%
e^{\beta \left( m-\delta \mu \right) }+1}\left[ \frac{1-q^{2}}{12}+q\alpha
_{0}\left( q\alpha _{0}-s\delta \right) \right] ,  \label{DelQa}
\end{equation}%
with $\Delta Q_{0}=se\alpha _{0}/2$. For $|\mu |<m$, in the zero temperature
limit one has $\lim_{T\rightarrow 0}\Delta Q=\Delta Q_{0}$ and the
topological part of the charge coincides with that for the vacuum state.

In the case $|\mu |>m$, the charge at zero temperature differs from the
vacuum charge $\Delta Q_{0}$. It is obtained by the integration of the
right-hand side of (\ref{j0T0}), omitting the term $l=0$ (the Minkowskian
part). By taking into account that
\begin{equation}
\int_{0}^{\infty }dy\,yg_{l}(ay)=\frac{1}{4a^{2}},\;l=1,2,  \label{Int4}
\end{equation}%
and using the relation (\ref{Rel4}), one finds the following expression%
\begin{equation}
\Delta Q_{T=0}=\Delta Q_{0}+\mathrm{sgn}(\mu )\frac{e}{2q}\left\{ \frac{%
1-q^{2}}{12}+q\alpha _{0}\left[ q\alpha _{0}-s\mathrm{sgn}(\mu )\right]
\right\} ,  \label{DelQT0b}
\end{equation}%
with the same $\Delta Q_{0}$ as in (\ref{DelQa}). For a given sign of the
chemical potential, the zero temperature charge is completely determined by
the topological parameters of the model. Note that in the evaluation of the
integral (\ref{Int4}) with the function $g_{1}(u)$ from (\ref{g12}) one
cannot change the order of integrations. In order to escape this difficulty,
we introduce in the integrand the function $e^{-by}$, $b>0$. With this
function, changing the integrations order, the integral over $y$ involving
the Bessel function is evaluated by using the formula from \cite{Prud86}.
Then, after the evaluation of the integral over $x$, we take the limit $%
b\rightarrow 0$. For $|\mu |=m$ the expression (\ref{DelQT0b}) coincides
with that obtained from (\ref{DelQa}) in the limit $T\rightarrow 0$.

\section{Current density}

\label{sec:Current}

In this section we consider the expectation value of the spatial components
for the current density. First of all we can see that the current density
along the radial direction is zero, $\langle j^{1}\rangle =0$, and the only
nonzero component is along the azimuthal direction. The contributions to the
azimuthal current coming from the particles and antiparticles are obtained
from (\ref{current4}) by using the mode functions (\ref{Modes}). For the
physical component, $\langle j_{\phi }\rangle $, connected to the
contravariant one by the relation $\langle j_{\phi }\rangle =r\langle
j^{2}\rangle $, one gets the following expression:%
\begin{equation}
\langle j_{\phi }\rangle _{\pm }=\pm \frac{qe}{2\pi }\sum_{j}\epsilon
_{j}\int_{0}^{\infty }d\gamma \,\frac{\gamma ^{2}}{E}\frac{J_{\beta
_{j}}(\gamma r)J_{\beta _{j}+\epsilon _{j}}(\gamma r)}{e^{\beta (E\mp \mu
)}+1}.  \label{j2pm}
\end{equation}%
Note that the expectation value of the azimuthal current density does not
depend on the parameter $s$ in (\ref{Dirac}) and it is the same for both the
irreducible representations of the Clifford algebra.

By making use of the recurrence relation for the Bessel function we can show
that%
\begin{equation}
\epsilon _{j}J_{\beta _{j}}(x)J_{\beta _{j}+\epsilon _{j}}(x)=\frac{1}{x}%
\left( \epsilon _{j}\beta _{j}-\frac{1}{2}x\partial _{x}\right) J_{\beta
_{j}}^{2}(x).  \label{Bess2}
\end{equation}%
Substituting this into (\ref{j2pm}) and using the representation (\ref{IntJ}%
), the contributions to the azimuthal current from the particles and
antiparticles are presented as%
\begin{equation}
\langle j_{\phi }\rangle _{\pm }=\frac{eqr}{2\pi }\int_{0}^{\infty }d\gamma
\,\frac{\gamma ^{3}/E}{e^{\beta (E\mp \mu )}+1}\frac{1}{2\pi i}%
\int_{c-i\infty }^{c+i\infty }\frac{dt}{t^{2}}e^{t/2-\gamma
^{2}r^{2}/t}\sum_{\delta =\pm 1}\delta \mathcal{I}(q,\delta \alpha
_{0},\gamma ^{2}r^{2}/t).  \label{j2pmn1}
\end{equation}%
By taking into account (\ref{Ic1}), the integral over $t$ is expressed in
terms of the Bessel function of the order 1 and we get%
\begin{eqnarray}
\langle j_{\phi }\rangle _{\pm } &=&\frac{e}{\pi }\int_{0}^{\infty }d\gamma
\,\frac{\gamma ^{2}/E}{e^{\beta (E\mp \mu )}+1}\left\{
\sum_{l=1}^{p}(-1)^{l}\sin \left( 2\pi l\alpha _{0}\right) J_{1}(2\gamma
rs_{l})\right.  \notag \\
&&\left. -\frac{q}{\pi }\int_{0}^{\infty }dy\,\frac{f_{2}(q,\alpha
_{0},y)J_{1}(2\gamma r\cosh y)}{\left[ \cosh (2qy)-\cos (q\pi )\right] \cosh
y}\right\} .  \label{j2pmn2}
\end{eqnarray}%
As is seen, the azimuthal current is an odd function of the parameter $%
\alpha _{0}$. Note that for the zero chemical potential, $\mu =0$, the
particles and antiparticles give the same contributions to the total current
density. As before, we will consider the cases $|\mu |\leqslant m$ and $|\mu
|>m$ separately.

For $|\mu |\leqslant m$, by using the expansion (\ref{Expansion}), one finds%
\begin{eqnarray}
\langle j_{\phi }\rangle _{\pm } &=&-\frac{2^{3/2}em^{3}r}{\pi ^{3/2}}%
\sum_{n=1}^{\infty }(-1)^{n}e^{\pm n\beta \mu }\left[
\sum_{l=1}^{p}(-1)^{l}s_{l}\sin (2\pi l\alpha _{0})\right.  \notag \\
&&\times \left. f_{3/2}\left( c_{nl}\right) -\frac{q}{\pi }\int_{0}^{\infty
}dy\,\frac{f_{2}(q,\alpha _{0},y)f_{3/2}\left( c_{n}(y)\right) }{\cosh
(2qy)-\cos (q\pi )}\right] .  \label{j2pm4}
\end{eqnarray}%
For the total current density this gives%
\begin{eqnarray}
\langle j_{\phi }\rangle &=&-\frac{2^{5/2}em^{3}r}{\pi ^{3/2}}%
\sideset{}{'}{\sum}_{n=0}^{\infty }(-1)^{n}\cosh (n\beta \mu )\left[
\sum_{l=1}^{p}(-1)^{l}f_{3/2}\left( c_{nl}\right) \right.  \notag \\
&&\left. \times s_{l}\sin (2\pi l\alpha _{0})-\frac{q}{\pi }\int_{0}^{\infty
}dy\,\frac{f_{2}(q,\alpha _{0},y)f_{3/2}\left( c_{n}(y)\right) }{\cosh
(2qy)-\cos (q\pi )}\right] ,  \label{j2}
\end{eqnarray}%
where the $n=0$ term corresponds to the vacuum expectation value \cite%
{Beze10}:%
\begin{equation}
\langle j_{\phi }\rangle _{0}=-\frac{2^{3/2}em^{3}r}{\pi ^{3/2}}\left[
\sum_{l=1}^{p}(-1)^{l}s_{l}\sin (2\pi l\alpha _{0})f_{3/2}\left(
2mrs_{l}\right) -\frac{q}{\pi }\int_{0}^{\infty }dy\,\frac{f_{2}(q,\alpha
_{0},y)f_{3/2}\left( 2mr\cosh y\right) }{\cosh (2qy)-\cos (q\pi )}\right] .
\label{j20}
\end{equation}%
Unlike to the FC and charge density, the azimuthal current density has
definite parity with respect to the reflections $\alpha _{0}\rightarrow
-\alpha _{0}$ and $\mu \rightarrow -\mu $: it is an odd function of $\alpha
_{0}$ and an even function of the chemical potential. For an integer $q$ and
for special values of $\alpha _{0}$, given by (\ref{alfsp}), the integral
terms in (\ref{j2}) and (\ref{j20}) vanish.

\begin{figure}[tbph]
\begin{center}
\epsfig{figure=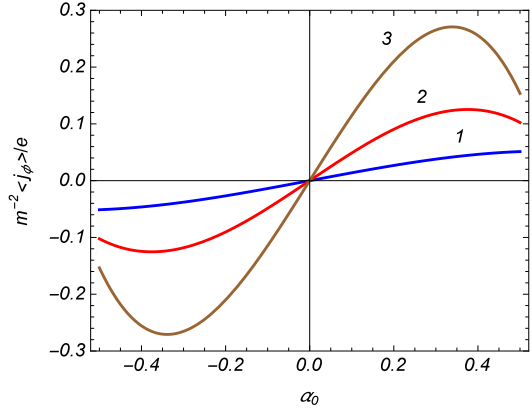,width=7.cm,height=5.5cm}
\end{center}
\caption{Azimuthal current density as a function of $\protect\alpha _{0}$
for separate values of the parameter $q$ (numbers near the curves) and for $%
\protect\mu /m=0.25$, $mr=0.5$, $T/m=0.5$.}
\label{fig4}
\end{figure}

In Figure \ref{fig4}, the azimuthal current density is plotted versus the
parameter $\alpha _{0}$ for $\mu /m=0.25$, $mr=0.5$, $T/m=0.5$. The numbers
near the curves are the values of $q$. As we have mentioned, the current
density is an odd function of $\alpha _{0}$.

For the massless field, due to the condition $\mu \leqslant m$, we also
should take $\mu =0$. Using the asymptotic expression for the Macdonald
function for small values of the argument, we find
\begin{eqnarray}
\langle j_{\phi }\rangle &=&-\frac{4er}{\pi }\sideset{}{'}{\sum}%
_{n=0}^{\infty }(-1)^{n}\left[ \sum_{l=1}^{p}\frac{(-1)^{l}s_{l}\sin (2\pi
l\alpha _{0})}{\left( n^{2}\beta ^{2}+4s_{l}^{2}r^{2}\right) ^{3/2}}\right.
\notag \\
&&\left. -\frac{q}{\pi }\int_{0}^{\infty }dy\,\frac{\left( n^{2}\beta
^{2}+4r^{2}\cosh ^{2}y\right) ^{-3/2}}{\cosh (2qy)-\cos (q\pi )}%
f_{2}(q,\alpha _{0},y)\right] .  \label{j2m0}
\end{eqnarray}%
In the case of the Minkwoski bulk with a magnetic flux, the current density
is obtained from (\ref{j2}) with $q=1$:%
\begin{equation}
\langle j_{\phi }\rangle =\frac{2^{5/2}em^{3}r}{\pi ^{5/2}}\sin (\pi \alpha
_{0})\sideset{}{'}{\sum}_{n=0}^{\infty }(-1)^{n}\cosh (n\beta \mu
)\int_{0}^{\infty }dy\cosh (2\alpha _{0}y)f_{3/2}\left( c_{n}(y)\right) .
\label{j2q1}
\end{equation}

At the cone apex, the thermal part in the azimuthal current density vanishes
as $r$ for $2|\alpha _{0}|<1-1/q$ and as $r^{2\rho }$ for $2|\alpha
_{0}|>1-1/q$. In the first case the leading term is obtained directly from
the $n\neq 0$ term in (\ref{j2}) putting $r=0$. This is reduced to the
substitutions $c_{nl}=c_{n}(y)=mn\beta $. In the second case, the leading
term in the asymptotic expansion of the thermal part is found in the way
similar to that we have used for the FC and has the form%
\begin{equation}
\langle j_{\phi }\rangle \approx \langle j_{\phi }\rangle _{0}+\frac{%
em^{2}q(mr)^{2\rho }}{2^{\rho -1}\pi ^{5/2}}\Gamma (1/2-\rho )\mathrm{sgn}%
(\alpha _{0})\cos (\pi \rho )\sideset{}{'}{\sum}_{n=0}^{\infty
}(-1)^{n}\cosh (n\beta \mu )f_{\rho +1}(mn\beta ).  \label{j2r0}
\end{equation}%
The vacuum expectation value diverges as $1/r^{2}$ and it dominates near the
apex.

Considering the limit of low temperatures for a fixed value of $mr$, the
main contribution for the thermal part of the azimuthal current comes from
the $n=1$ term and the leading term is given by%
\begin{equation}
\langle j_{\phi }\rangle \approx \langle j_{\phi }\rangle _{0}+\frac{%
2emT^{2}r}{\pi e^{\beta (m-|\mu |)}}\left[ \sum_{l=1}^{p}(-1)^{l}s_{l}\sin
(2\pi l\alpha _{0})-\frac{q}{\pi }\int_{0}^{\infty }dy\,\frac{f_{2}(q,\alpha
_{0},y)}{\cosh (2qy)-\cos (q\pi )}\right] .  \label{j2smT}
\end{equation}%
In this limit the contribution of the finite temperature effects is
suppressed by the factor $e^{-\beta (m-|\mu |)}$.

An alternative representation for the current density in the case $|\mu
|\leqslant m$ is obtained by using the formula (\ref{rep}):%
\begin{eqnarray}
\langle j_{\phi }\rangle &=&-\frac{2eT}{\pi }\sum_{n=-\infty }^{+\infty
}b_{n}\left\{ \sum_{l=1}^{p}(-1)^{l}\sin (2\pi l\alpha _{0})K_{1}\left(
2rs_{l}b_{n}\right) \right.  \notag \\
&&\left. -\frac{q}{\pi }\int_{0}^{\infty }dy\,\frac{f_{2}(q,\alpha
_{0},y)K_{1}\left( 2rb_{n}\cosh y\right) }{[\cosh (2qy)-\cos (q\pi )]\cosh y}%
\right\} ,  \label{j2alt}
\end{eqnarray}%
where $b_{n}$ is given by (\ref{bn}).

At high temperatures, assuming that $rT\gg 1$, the dominant contribution in (%
\ref{j2alt}) comes from the terms $n=0$ and $n=-1$. For $q>2$ one finds%
\begin{equation}
\langle j_{\phi }\rangle \approx \frac{2eT^{3/2}}{\sqrt{s_{1}r}}\sin (2\pi
\alpha _{0})\frac{\cos (2rs_{1}\mu )}{e^{2\pi rs_{1}T}}.  \label{j2highT}
\end{equation}%
In the case $q<2$, assuming that $T\gg 1/[\pi r\sin ^{2}(q\pi /2)]$, the
leading term is given by%
\begin{equation}
\langle j_{\phi }\rangle \approx \frac{eqT\sin (\pi q\alpha _{0})\cos (2r\mu
)}{\pi r\sin (\pi q/2)e^{2\pi rT}}.  \label{j2highTb}
\end{equation}

\begin{figure}[tbph]
\begin{center}
\epsfig{figure=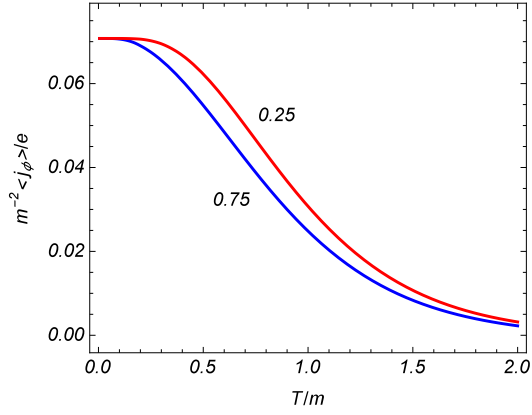,width=7.cm,height=5.5cm}
\end{center}
\caption{Azimuthal current density as a function of temperature. The numbers
near the curves are the values of the ratio $\protect\mu /m$ and the graphs
are plotted for $q=1.5$, $\protect\alpha _{0}=0.25$, $mr=0.5$.}
\label{fig5}
\end{figure}

The dependence of the current density on temperature is plotted in Figure %
\ref{fig5} for two values of the ratio $\mu /m$ (numbers near the graphs).
The graphs are plotted for $q=1.5$, $\alpha _{0}=0.25$, $mr=0.5$.

For the investigation of the asymptotic of the azimuthal current al large
distances from the origin we again use the representation (\ref{j2alt}). In
the case $q>2$, the dominant contribution comes from the terms with $l=1$, $%
n=0,1$ and we get%
\begin{equation}
\langle j_{\phi }\rangle \approx \frac{2eT\sin (2\pi \alpha _{0})}{\sqrt{\pi
s_{1}r}}\mathrm{Re\,}\left( b_{0}^{1/2}e^{-2rs_{1}b_{0}}\right) .
\label{j2larger}
\end{equation}%
For $q<2$ and not too close to 2, the leading term is given by the expression%
\begin{equation}
\langle j_{\phi }\rangle \approx \frac{eq\sin (\pi q\alpha _{0})T}{\pi \sin
(\pi q/2)r}\mathrm{Re\,}\left( e^{-2rb_{0}}\right) .  \label{j2largr2}
\end{equation}

Now let us consider the current density for the case $|\mu |>m$. The
corresponding expression has the form%
\begin{eqnarray}
\langle j_{\phi }\rangle &=&\langle j_{\phi }\rangle _{0}+\langle j_{\phi
}\rangle _{\pm }+\frac{e}{\pi }\int_{0}^{\infty }d\gamma \,\frac{\gamma
^{2}/E}{e^{\beta (E-|\mu |)}+1}\left\{ \sum_{l=1}^{p}(-1)^{l}\sin \left(
2\pi l\alpha _{0}\right) \right.  \notag \\
&&\times \left. J_{1}(2\gamma rs_{l})-\frac{q}{\pi }\int_{0}^{\infty }dy\,%
\frac{f_{2}(q,\alpha _{0},y)J_{1}(2\gamma r\cosh y)}{\left[ \cosh (2qy)-\cos
(q\pi )\right] \cosh y}\right\} ,  \label{j2v2}
\end{eqnarray}%
where the upper and lower signs correspond to the cases $\mu <-m$ and $\mu
>m $, respectively. The expression for $\langle j_{\phi }\rangle _{\pm }$ in
the right-hand side is given by (\ref{j2pm4}). At zero temperature, $%
T\rightarrow 0$, we get%
\begin{eqnarray}
\langle j_{\phi }\rangle _{T=0} &=&\langle j_{\phi }\rangle _{0}-\frac{%
ep_{0}^{3}}{2\pi m}\left\{ \sum_{l=1}^{p}(-1)^{l}\sin \left( 2\pi l\alpha
_{0}\right) g_{1}^{\prime }(p_{0}rs_{l})\right.  \notag \\
&&\left. -\frac{q}{\pi }\int_{0}^{\infty }dy\,\frac{f_{2}(q,\alpha
_{0},y)g_{1}^{\prime }(p_{0}r\cosh y)}{\left[ \cosh (2qy)-\cos (q\pi )\right]
\cosh y}\right\} ,  \label{j2T0}
\end{eqnarray}%
with the function $g_{1}(u)$ from (\ref{g12}) and $g_{1}^{\prime
}(u)=\partial _{u}g_{1}(u)$. The second term in the right-hand side is the
contribution from the antiparticles for $\mu <-m$ and from the particles for
$\mu >m$. In the case of a massless field we get
\begin{equation}
\langle j_{\phi }\rangle _{T=0}=\frac{e}{4\pi r^{2}}\left\{ \sum_{l=1}^{p}%
\frac{(-1)^{l}}{s_{l}^{2}}\sin \left( 2\pi l\alpha _{0}\right) g_{0}(|\mu
|rs_{l})-\frac{q}{\pi }\int_{0}^{\infty }dy\,\frac{f_{2}(q,\alpha
_{0},y)g_{0}(|\mu |r\cosh y)}{\left[ \cosh (2qy)-\cos (q\pi )\right] \cosh
^{3}y}\right\} ,  \label{j2T0m0}
\end{equation}%
with the function
\begin{equation}
g_{0}(u)=\int_{0}^{2u}dxxJ_{1}(x)-1.  \label{g0u}
\end{equation}%
The part in (\ref{j2T0m0}) coming from $-1$ in the right-hand side of (\ref%
{g0u}) corresponds to the vacuum current density. It depends on the radial
coordinate as $1/r^{2}$.

\begin{figure}[tbph]
\begin{center}
\begin{tabular}{cc}
\epsfig{figure=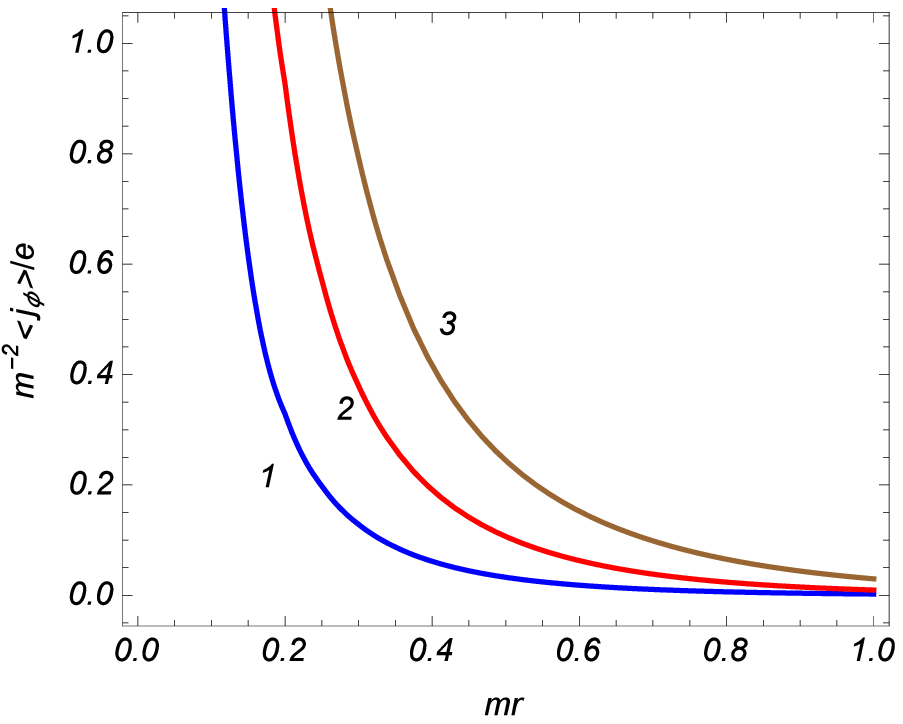,width=7.cm,height=5.5cm} & \quad %
\epsfig{figure=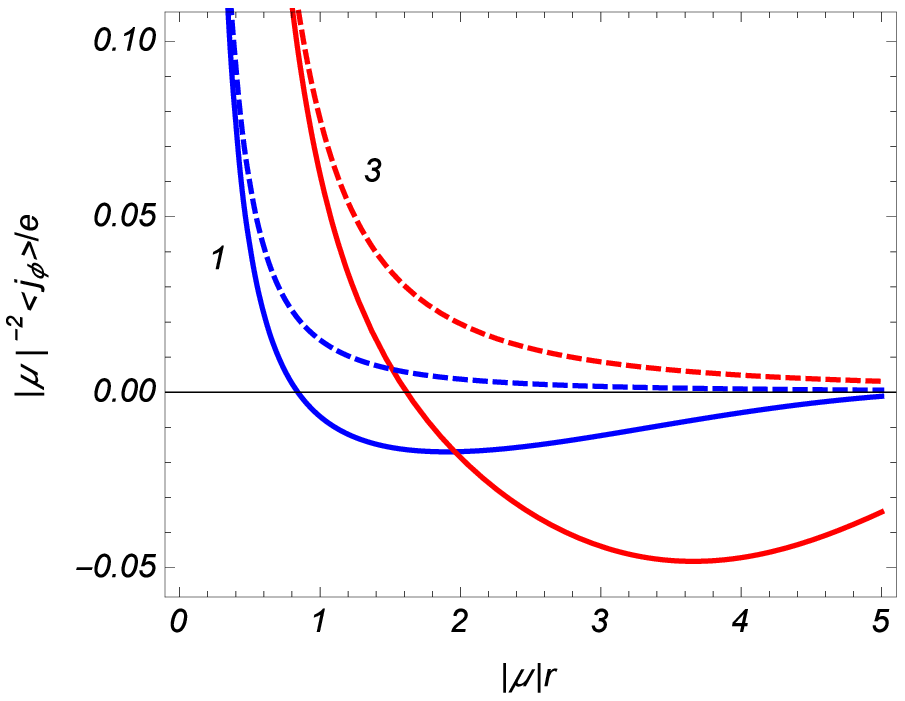,width=7.cm,height=5.5cm}%
\end{tabular}%
\end{center}
\caption{Current density versus the distance from the apex for $\protect%
\alpha _{0}=0.25$ and for different values of the parameter $q$ (the numbers
near the graphs). The graphs on the left panel are plotted for $\protect\mu %
/m=0.25$, $T/m=0.5$. The right panel presents the zero temperature (full
curves) and vacuum currents ( dashed curves) for a massless field. }
\label{fig6}
\end{figure}

The dependence of the azimuthal current density on the distance from the
cone apex is displayed in Figure \ref{fig6} for $\alpha _{0}=0.25$ and for
separate values of $q$ (numbers near the curves). The graphs on the left
panel are plotted for $\mu /m=0.25$, $T/m=0.5$. The full curves on the right
panel present the current density at zero temperature for a massless
fermionic field with the chemical potential $\mu $. The dashed curves are
the vacuum current densities in the same model.

\section{Expectation values in parity and time-reversal symmetric models}

\label{sec:Tsym}

In the discussion above we have considered a fermionic field realizing the
irreducible representation of the Clifford algebra. In this representation,
the mass term in the Lagrangian density is not invariant under the parity
and time-reversal transformations ($P$- and $T$-transformations). In order
to recover the $P$- and $T$-invariance, we note that in (2+1)-dimensions the
$\gamma ^{2}$ matrix can be represented in two different ways: $\gamma
^{2}=\gamma _{(s)}^{2}=-is\sqrt{-g^{22}}\gamma ^{0}\gamma ^{1}$, with $s=\pm
1$. As a consequence, the Clifford algebra has two inequivalent
representations corresponding to the upper and lower signs. Our choice in (%
\ref{gamma}) corresponds to the representation with the upper sign. Consider
two two-component spinor fields, $\psi _{(+1)}$ and $\psi _{(-1)}$, with the
combined Lagrangian density $\mathcal{L}=\sum_{s=\pm 1}\bar{\psi}%
_{(s)}(i\gamma _{(s)}^{\mu }D_{\mu }-m)\psi _{(s)}$, where $\gamma
_{(s)}^{\mu }=(\gamma ^{0},\gamma ^{1},\gamma _{(s)}^{2})$. We can see that,
by suitable transformations of the fields (see, for example, \cite{Shim85}),
this Lagrangian is invariant under $P$- and $T$-transformations (in the
absence of magnetic fields). Defining new fields $\psi _{(+1)}^{\prime
}=\psi _{(+1)}$, $\psi _{(-1)}^{\prime }=\gamma ^{0}\gamma ^{1}\psi _{(-1)}$%
, the Lagrangian density is transformed to the form%
\begin{equation}
\mathcal{L}=\sum_{s=\pm 1}\bar{\psi}_{(s)}^{\prime }(i\gamma ^{\mu }D_{\mu
}-sm)\psi _{(s)}^{\prime },  \label{Lag2}
\end{equation}%
with $\gamma ^{\mu }=(\gamma ^{0},\gamma ^{1},\gamma _{(+1)}^{2})$. From
here it follows that the field $\psi _{(-1)}^{\prime }$ satisfies the same
equation as $\psi _{(+1)}$ with the opposite sign for the mass term. We can
write the Lagrangian (\ref{Lag2}) in terms of the four-component spinor $%
\Psi =(\psi _{(+1)}^{\prime },\psi _{(-1)}^{\prime })^{T}$ as $\mathcal{L}=%
\bar{\Psi}(i\gamma _{(1)}^{\mu }D_{\mu }-m\eta )\Psi $ with $4\times 4$
matrices $\gamma _{(1)}^{\mu }=I_{2}\otimes \gamma ^{\mu }$ and $\eta
=\sigma _{3}\otimes I_{2}$, where $I_{2}=\mathrm{diag}(1,1)$. Another form
is obtained by using the $4\times 4$ reducible representation of gamma
matrices $\gamma _{(2)}^{\mu }=\sigma _{3}\otimes \gamma ^{\mu }$ with the
Lagrangian density $\mathcal{L}=\bar{\Psi}(i\gamma _{(2)}^{\mu }D_{\mu
}-m)\Psi $.

As is seen from (\ref{Lag2}), the expectation values in the corresponding
models are obtained from the formulas given above by summing the
contributions from the fields $\psi _{(+1)}$ and $\psi _{(-1)}$. For the
first one the expressions are obtained from those presented in the previous
sections taking $s=1$. In order to find the contribution of the field $\psi
_{(-1)}$, we note that $\bar{\psi}_{(-1)}\psi _{(-1)}=-\bar{\psi}%
_{(-1)}^{\prime }\psi _{(-1)}^{\prime }$ and $\bar{\psi}_{(-1)}\gamma
_{(-1)}^{\mu }\psi _{(-1)}=\bar{\psi}_{(-1)}^{\prime }\gamma ^{\mu }\psi
_{(-1)}^{\prime }$, with $\mu =0,2$. The expectation values $\langle \bar{%
\psi}_{(-1)}^{\prime }\psi _{(-1)}^{\prime }\rangle $ and $\langle \bar{\psi}%
_{(-1)}^{\prime }\gamma ^{\mu }\psi _{(-1)}^{\prime }\rangle $ are obtained
from the formulas in the previous sections taking $s=-1$. As a result, for
the total expectation values we get%
\begin{eqnarray}
\langle \bar{\psi}\psi \rangle &=&\sum_{s=\pm 1}\langle \bar{\psi}_{(s)}\psi
_{(s)}\rangle =\sum_{s=\pm 1}s\langle \bar{\psi}_{(s)}^{\prime }\psi
_{(s)}^{\prime }\rangle ,  \notag \\
\langle \bar{\psi}\gamma ^{\mu }\psi \rangle &=&\sum_{s=\pm 1}\langle \bar{%
\psi}_{(s)}\gamma _{(s)}^{\mu }\psi _{(s)}\rangle =\sum_{s=\pm 1}\langle
\bar{\psi}_{(s)}^{\prime }\gamma ^{\mu }\psi _{(s)}^{\prime }\rangle .
\label{tot}
\end{eqnarray}

Here we give the final expressions. First of all, note that the
contributions to the azimuthal current density from the fields $\psi _{(+1)}$
and $\psi _{(-1)}$ coincide and the expression for the total current is
obtained from (\ref{j2}) with the additional coefficient 2.

For the FC, summing the separate contributions in accordance with (\ref{tot}%
), in the case $|\mu |\leqslant m$ one has%
\begin{eqnarray}
\langle \bar{\psi}\psi \rangle &=&\langle \bar{\psi}\psi \rangle ^{(M)}-%
\frac{2^{5/2}m^{2}}{\pi ^{3/2}}\sideset{}{'}{\sum}_{n=0}^{\infty
}(-1)^{n}\cosh (n\beta \mu )\left[ \sum_{l=1}^{p}(-1)^{l}\right.  \notag \\
&&\times \left. c_{l}\cos (2\pi l\alpha _{0})f_{1/2}(c_{nl})-\frac{q}{\pi }%
\int_{0}^{\infty }dy\,\frac{f_{1}(q,\alpha _{0},y)f_{1/2}(c_{n}(y))}{\cosh
(2qy)-\cos (q\pi )}\right] ,  \label{FCsym}
\end{eqnarray}%
where the contribution $\langle \bar{\psi}\psi \rangle ^{(M)}$ is given by (%
\ref{FCM}) with the additional coefficient 2 and taking $s=1$. The FC in the
vacuum state, $\langle \bar{\psi}\psi \rangle _{0}$, corresponds to the $n=0$
term in (\ref{FCsym}). Now, the FC is an even function of the chemical
potential $\mu $ and of the magnetic flux parameter $\alpha _{0}$. For $|\mu
|>m$, the FC at zero temperature contains two contributions:%
\begin{eqnarray}
\langle \bar{\psi}\psi \rangle _{T=0} &=&\langle \bar{\psi}\psi \rangle _{0}+%
\frac{2p_{0}^{2}}{\pi }\left[ \sideset{}{'}{\sum}_{l=0}^{p}(-1)^{l}c_{l}\cos
(2\pi l\alpha _{0})g_{1}(p_{0}rs_{l})\right.  \notag \\
&&\left. -\frac{q}{\pi }\int_{0}^{\infty }dy\,\frac{f_{1}(q,\alpha
_{0},y)g_{1}(p_{0}r\cosh y)}{\cosh (2qy)-\cos (q\pi )}\right] .
\label{FCsymT0}
\end{eqnarray}%
The second term comes from the particles for $\mu >0$ and antiparticles for $%
\mu <0$. It is symmetric under the change of the sign for the chemical
potential, $\mu \rightarrow -\mu $.

\begin{figure}[tbph]
\begin{center}
\epsfig{figure=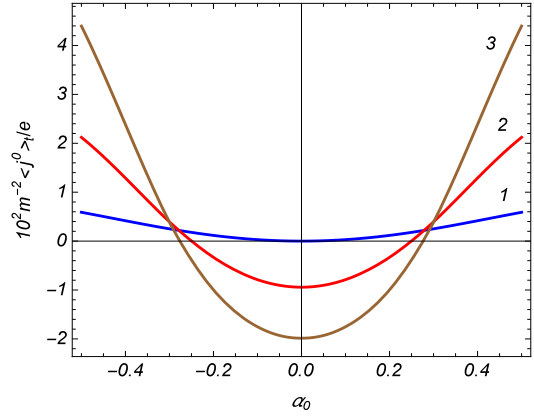,width=7.cm,height=5.5cm}
\end{center}
\caption{The same as in Figure \protect\ref{fig1} in a $P$- and $T$%
-invariant model with two spinors $\protect\psi _{(+1)}$ and $\protect\psi %
_{(-1)}$. }
\label{fig7}
\end{figure}

For the charge density in $P$- and $T$-symmetric models, in the case $|\mu
|\leqslant m$, we get%
\begin{eqnarray}
\langle j^{0}\rangle &=&\langle j^{0}\rangle ^{(M)}-\frac{2^{5/2}em^{3}}{\pi
^{3/2}T}\sum_{n=1}^{\infty }(-1)^{n}n\sinh {(n\beta \mu )}\left[
\sum_{l=1}^{p}(-1)^{l}c_{l}\right.  \notag \\
&&\times \left. \cos (2\pi l\alpha _{0})f_{3/2}\left( c_{nl}\right) -\frac{q%
}{\pi }\int_{0}^{\infty }dy\,\frac{f_{1}(q,\alpha _{0},y)f_{3/2}\left(
c_{n}(y)\right) }{\cosh (2qy)-\cos (q\pi )}\right] ,  \label{j0sym}
\end{eqnarray}%
with the Minkowskian part from (\ref{j0M}) with the additional coefficient
2. Note that the vacuum charge density vanishes. The charge density is an
even function of $\alpha _{0}$ and an odd function of the chemical
potential. In particular, it vanishes for the zero chemical potential. This
is because of the cancellation of the contributions coming from the separate
irreducible representations. In Figure \ref{fig7}, the topological part in
the charge density is displayed as a function of $\alpha _{0}$ in the $P$-
and $T$-invariant model with two spinors $\psi _{(+1)}$ and $\psi _{(-1)}$.
The numbers near the curves are the values of the parameter $q$ and we have
taken $\mu /m=0.25$, $mr=0.5$, $T/m=0.5$.

\begin{figure}[tbph]
\begin{center}
\epsfig{figure=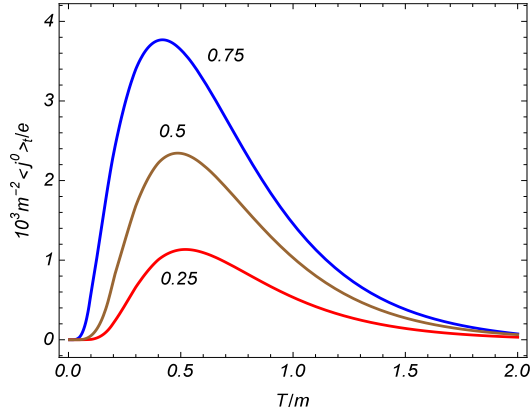,width=7.cm,height=5.5cm}
\end{center}
\caption{The same as in Figure \protect\ref{fig2} in a $P$- and $T$%
-invariant model. The numbers near the curves are the values of the ratio $%
\protect\mu /m$.}
\label{fig8}
\end{figure}

The dependence of the topological part in the charge density on the
temperature is presented in Figure \ref{fig8} for different values of the
ratio $\mu /m$ (numbers near the curves). For the other parameters we have
taken the same values as in Figure \ref{fig2}.

In the case $|\mu |>m$, the charge density is obtained by summing the
expressions in the right-hand side of (\ref{j0v2}) for $s=1$ and $s=-1$. In
particular, at zero temperature we have%
\begin{eqnarray}
\langle j^{0}\rangle _{T=0} &=&\frac{2}{\pi }\mathrm{sgn}(\mu )ep_{0}^{2}%
\left[ \sideset{}{'}{\sum}_{l=0}^{p}(-1)^{l}c_{l}\cos (2\pi l\alpha
_{0})g_{2}(p_{0}rs_{l})\right.  \notag \\
&&\left. -\frac{q}{\pi }\int_{0}^{\infty }dy\,\frac{f_{1}(q,\alpha
_{0},y)g_{2}(p_{0}r\cosh y)}{\cosh (2qy)-\cos (q\pi )}\right] .
\label{j0symT0}
\end{eqnarray}%
The zero temperature charge density comes from the particles or
antiparticles for the cases $\mu >0$ and $\mu <0$ respectively.

\begin{figure}[tbph]
\begin{center}
\begin{tabular}{cc}
\epsfig{figure=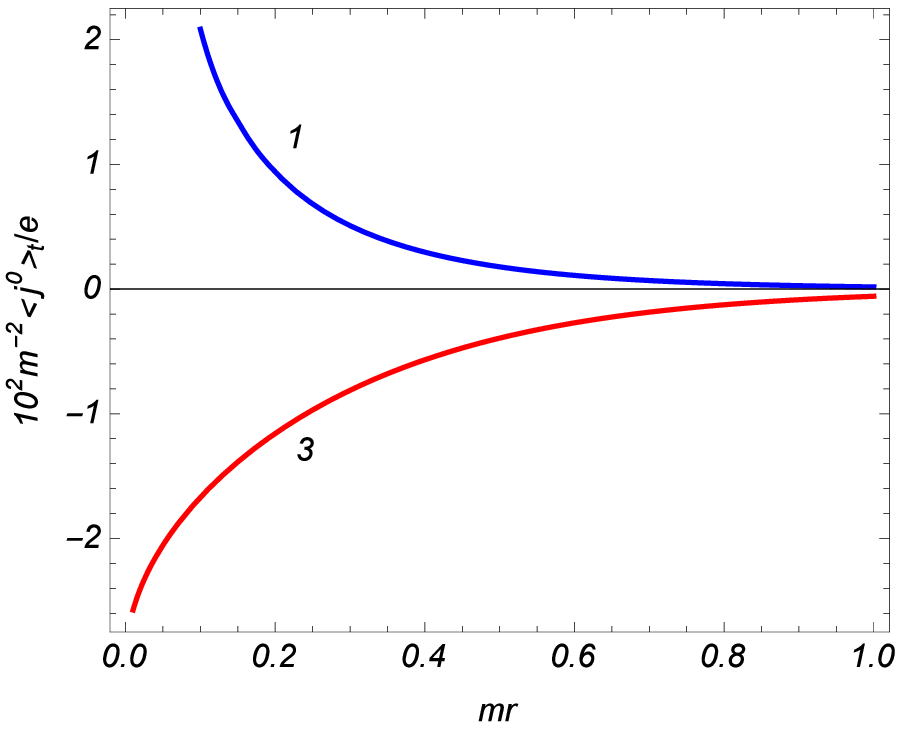,width=7.cm,height=5.5cm} & \quad %
\epsfig{figure=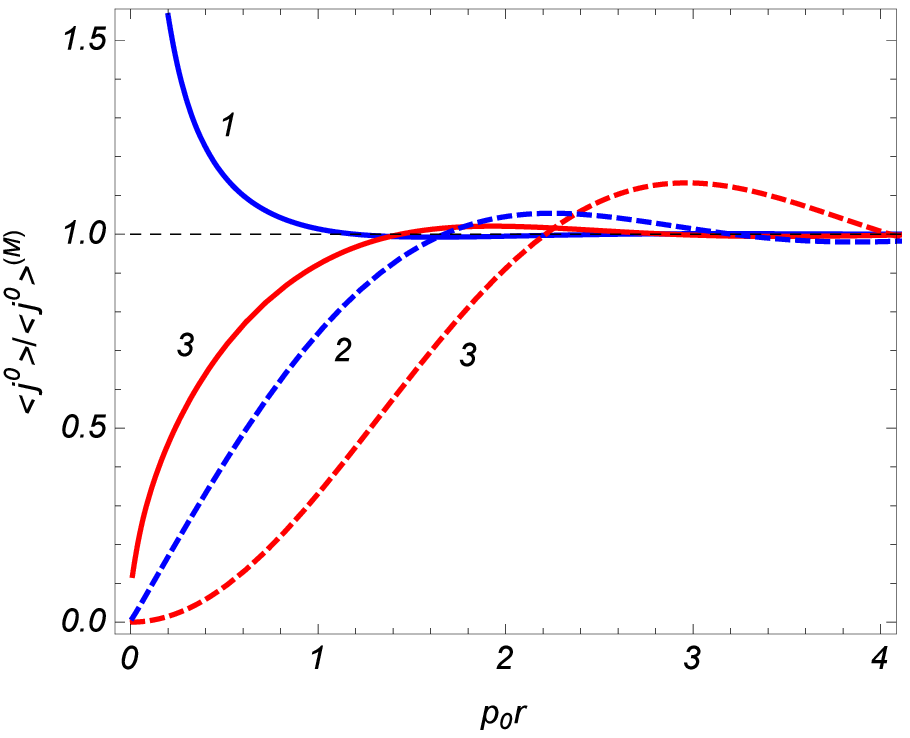,width=7.cm,height=5.5cm}%
\end{tabular}%
\end{center}
\caption{Charge density in the $P$- and $T$-invariant model as a function of
the distance from the apex for separate values of $q$ (the numbers near the
curves). On the left panel, the topological part is presented for $\protect%
\mu /m=0.25$, $\protect\alpha _{0}=0.25$, $T/m=0.5$. The right panel
displays the ratio $\langle j^{0}\rangle /\langle j^{0}\rangle ^{(M)}$ at
zero temperature. The full and dashed curves correspond to $\protect\alpha %
_{0}=0.25$ and $\protect\alpha _{0}=0$.}
\label{fig9}
\end{figure}

In Figure \ref{fig9} the dependence of the charge density on the radial
coordinate is plotted for different values of $q$ (numbers near the graphs).
For the left panel we have taken $\mu /m=0.25$, $\alpha _{0}=0.25$, $T/m=0.5$%
. Note that for these values of the parameters and in the case $q=2$ the
topological part vanishes. On the right panel the ratio $\langle
j^{0}\rangle /\langle j^{0}\rangle ^{(M)}$ is plotted at zero temperature.
The full and dashed curves correspond to $\alpha _{0}=0.25$ and $\alpha
_{0}=0$, respectively. For $\alpha _{0}=0.25$ and $q=2$ one has $\langle
j^{0}\rangle =\langle j^{0}\rangle ^{(M)}$.

For the chemical potential in the range $|\mu |\leqslant m$, the total
charge induced by the planar angle deficit and by the magnetic flux is
presented as%
\begin{equation}
\Delta Q=eq\left( \frac{1/q^{2}-1}{12}+\alpha _{0}^{2}\right) \sum_{\delta
=\pm 1}\frac{\delta }{e^{\beta \left( m-\delta \mu \right) }+1}.
\label{DelQsym}
\end{equation}%
In the case of zero chemical potential the charge vanishes. For $|\mu
|\geqslant m$, for the topological part in the charge at zero temperature
one gets
\begin{equation}
\Delta Q_{T=0}=\mathrm{sgn}(\mu )eq\left( \frac{1/q^{2}-1}{12}+\alpha
_{0}^{2}\right) .  \label{DelQsymb}
\end{equation}%
This charge is completely determined by the topological parameters of the
model.

In summing the contributions from the spinors $\psi _{(+1)}$ and $\psi
_{(-1)}$ we have assumed that the parameter $\alpha $ is the same for both
these spinors. However, in general, this parameter can be different for
separate spinors. In particular, this difference can be a consequence of
different phases in the quasiperiodicity condition (\ref{periodcond}) along
the azimuthal direction.

Among the most important physical systems, with the low-energy sector
described by a Dirac-like theory in two spatial dimensions, is graphene. The
long-wavelength excitations of the electronic subsystem in graphene are
described by a pair of two-component spinors $\psi _{+}$ and $\psi _{-}$,
corresponding to the two different inequivalent points $\mathbf{K}_{+}$ and $%
\mathbf{K}_{-}$ at the corners of the two-dimensional Brillouin zone (see
\cite{Gusy07}). The separate components of these spinors give the amplitude
of the wave function on the $A$ and $B$ triangular sublattices of the
graphene hexagonal lattice. Graphitic cones are obtained from planar
graphene sheet if one or more sectors with the angle $\pi /3$ are excised
and the remainder is joined. The opening angle of the cone is connected to
the number of the removed sectors, $N_{c}$, by the relation $\phi _{0}=2\pi
(1-N_{c}/6)$ , with $N_{c}=1,2,\ldots ,5$. All these angles have been
observed in experiments \cite{Kris97}. At the apex of the graphitic cone the
hexagon of the planar graphene lattice is replaced by a polygon having $%
6-N_{c}$ sites. The periodicity conditions for the combined bispinor $\Psi
=(\psi _{+},\psi _{-})$ under the rotation around the cone apex are
discussed in \cite{Lamm00}. For even values of $N_{c}$ these conditions do
not mix the spinors $\psi _{+}$ and $\psi _{-}$, and we can apply the
formulas given above.

\section{Conclusion}

\label{sec:Conc}

We have investigated the expectation values of the FC, charge and current
densities for a massive fermion field with nonzero chemical potential in
thermal equilibrium on the background of (2+1)-dimensional conical spacetime
with an arbitrary planar angle deficit in the presence of a magnetic flux
located at the cone apex. For both the spinor fields realizing the two
inequivalent representations of the Clifford algebra, the expectation values
are decomposed into three contributions coming from the vacuum expectation
values, from the particles and from the antiparticles. All these
contributions are periodic functions of the magnetic flux with the period
equal to the flux quantum. The vacuum expectation values have been
investigated earlier and here we are mainly concerned with the finite
temperature effects.

In the case $|\mu |\leqslant m$, the FC is presented in the form (\ref{FC}),
where $s=1$ and $s=-1$ correspond to two irreducible representations of the
Clifford algebra. With these representations, the mass term breaks the $P$-
and $T$-invariances and, related to this, the FC has no definite parity with
respect to the reflections $\alpha _{0}\rightarrow -\alpha _{0}$ and $\mu
\rightarrow -\mu $. For a massless field with the zero chemical potential
the FC vanishes. In the massive case, for integer values of the parameter $q$
and for special values of the magnetic flux given by (\ref{alfsp}), the
integral terms in (\ref{FC}) vanish. In the absence of the magnetic flux,
the FC is given by (\ref{FC12}) and it has opposite signs for two
irreducible representations with the same modulus. Another special case
corresponds to Minkowski bulk in the presence of a magnetic flux ($q=1$)
with the FC given in (\ref{FCq1}). In order to clarify the behavior of the
FC, we have considered different asymptotics of the general formula. The
finite temperature part in the FC is finite on the apex for $2|\alpha
_{0}|<1-1/q$ and diverges as $1/r^{1-2\rho }$, with $\rho =q\left(
1/2-|\alpha _{0}|\right) $, in the case $2|\alpha _{0}|>1-1/q$. The
divergence is related to the irregular mode. For a massive field, the vacuum
FC diverges on the apex as $1/r$ and it dominates for points near the
origin. At low temperatures and for $|\mu |<m$ the finite temperature
effects are suppressed by the factor $e^{-(m-|\mu |)/T}$. In order to
investigate the high temperature asymptotic, for the FC we have provided an
alternative representation (\ref{FC13}). At high temperatures, for points
not too close to the origin, the FC is dominated by the Minkowskian part.
The effects induced by the planar angle deficit and by the magnetic flux are
suppressed by the factor $e^{-2\pi rT\sin (\pi /q)}$ for $q>2$ and by the
factor $e^{-2\pi rT}$ for $q<2$. The asymptotics at large distance from the
cone apex are given by the expressions (\ref{FClarg1}) and (\ref{Fclarg2})
for the cases $q>2$ and $q<2$, respectively. The expression for the FC in
the case $|\mu |>m$ takes the form (\ref{FCv2}) with the upper and lower
signs corresponding to $\mu <-m$ and $\mu >m$. Now, the FC at zero
temperature, given by (\ref{FCT0}), in addition to the vacuum part contains
a contribution coming from the antiparticles ($\mu <-m$) or particles ($\mu
>m$) filling the states with the energies $m\leqslant E\leqslant |\mu |$.
For points near the apex, the zero temperature FC is dominated by the vacuum
part whereas at large distances the contributions from particles or
antiparticles dominate.

The contributions from particles and antiparticles to the charge density are
given by (\ref{j0pm12}) and are further transformed to (\ref{j0pm1}) in the
case $|\mu |\leqslant m$. For the total charge density one has the
representation (\ref{j0}). In the case of a massless field with the zero
chemical potential, as a consequence of the cancellation of the
contributions from particles and antiparticles, the charge density vanishes.
Similar to the FC, the charge density has indefinite parity with respect to
the changes of the signs for $\alpha _{0}$ and $\mu $. In the absence of the
magnetic flux, the general expression is simplified to (\ref{j0flux0}) and
the charge density is an odd function of the chemical potential. The charge
density for another special case of Minkowski bulk with magnetic flux is
given by (\ref{j0q1}). The behavior of the thermal part in the charge
density near the apex is similar to that for the FC. In this region the
total charge density behaves as $1/r$ and is dominated by the vacuum part.
At large distances from the origin, the behavior of the topological part in
the charge density is given by (\ref{j0larger}) and (\ref{j0largr2}) for $%
q>2 $ and $q<2$, respectively. In this region one has an exponential
suppression of the topological contributions. At low temperatures and for $%
|\mu |<m$ the charge density is dominated by the vacuum part and the thermal
effects are suppressed by the factor $e^{-(m-|\mu |)/T}$. At high
temperatures, the main contribution comes from the Minkowskian part and the
topological part behaves as $e^{-2\pi rT\sin (\pi /q)}$ and $e^{-2\pi rT}$
in the cases $q>2$ and $q<2$, respectively. For the values of the chemical
potential $|\mu |>m$, the expression for the charge density takes the form (%
\ref{j0v2}) with the upper and lower signs corresponding to $\mu <-m$ and $%
\mu >m$. The contribution from the antiparticles or particles to zero
temperature charge density is given by the second term in the right-hand
side of (\ref{j0T0}). For a massless field this term survives only and the
expression is simplified to (\ref{j0T0m0}).

The total charge induced by the planar angle deficit and by the magnetic
flux is finite. For $|\mu |\leqslant m$ it is given by the expression (\ref%
{DelQa}) with $\Delta Q_{0}$ being the vacuum charge. In the case $|\mu |>m$%
, the charge at zero temperature receives an additional con tribution from
particles or antiparticles, depending on the sign of the chemical potential.
This contribution is given by the second term in the right-hand side of (\ref%
{DelQT0b}). For a given sign of the chemical potential it is completely
determined by the topological parameters of the model, $q$ and $\alpha _{0}$.

In the problem under consideration, the only nonzero component of the
current density is along the azimuthal direction. This component does not
depend on the representation of the Clifford algebra. For the chemical
potential in the region $|\mu |\leqslant m$, the corresponding expectation
value is presented as (\ref{j2}). The current density has definite parity
with respect to the reflections $\alpha _{0}\rightarrow -\alpha _{0}$ and $%
\mu \rightarrow -\mu $: it is an odd function of $\alpha _{0}$ and an even
function of $\mu $. In particular, the current density vanishes in the
absence of the magnetic flux. For a massless field the general expression is
simplified to (\ref{j2m0}) and in Minkowski bulk with the magnetic flux one
has the expression (\ref{j2q1}). The thermal part of the physical component
of the azimuthal current vanishes on the apex as $r$ for $2|\alpha
_{0}|<1-1/q$ and as $r^{2\rho }$ in the case $2|\alpha _{0}|>1-1/q$. The
vacuum current diverges as $1/r^{2}$ and dominates for points near the apex.
At low temperatures and for $|\mu |<m$ the finite temperature contribution
is given by the second term in the right-hand side of (\ref{j2r0}) with the
exponential suppression. At high temperatures the current density is
suppressed by the factor $e^{-2\pi rT\sin (\pi /q)}$ in the case $q>2$ and
by $e^{-2\pi rT}$ for $q<2$. The large distance asymptotics are given by (%
\ref{j2larger}) and (\ref{j2largr2}) in these two regions of $q$. For $|\mu
|>m$, the expression for the current density has the form (\ref{j2v2}) and
the zero temperature current density is given by (\ref{j2T0}). The latter
consists two parts: the vacuum current and the current from particles or
antiparticles filling the states with the energies $m\leqslant E\leqslant
|\mu |$.

One can construct parity and time-reversal symmetric (2+1)-dimensional
fermionic model by combining spinors realizing the two irreducible
representations of the Clifford algebra. The corresponding Lagrangian
density can be transformed to the form (\ref{Lag2}) with $s=\pm 1$. The
expectation values in this model are obtained by using the formulas for
separate representations (see (\ref{tot})). The resulting FC is an even
functions of both the chemical potential and the parameter $\alpha _{0}$.
The charge density is an odd function of the chemical potential and an even
function of $\alpha _{0}$. The vacuum charge density vanishes.

\section*{Acknowledgments}

E. R. B. M. thanks Conselho Nacional de Desenvolvimento Cient\'{\i}fico e
Tecnol\'{o}gico (CNPq), Process No. 313137/2014-5, for partial financial
support. E.B. thanks the Brazilian agency CAPES for the financial support.
A.A.S. was supported by the State Committee of Science Ministry of Education
and Science RA, within the frame of Grant No. SCS 15T-1C110, and by the
Armenian National Science and Education Fund (ANSEF) Grant No. hepth-4172.
The work was partially supported by GRAPHENE-Graphene-Based Revolutions in
ICT and Beyond, project n.604391 FP7-ICT-2013-FET-F, as well as by the NATO
Science for Peace Program under grant SFP 984537.

\appendix

\section{Appendix}

\label{sec:Append}

In this appendix we derive the formulas used in Section \ref{sec:Charge} for
the simplification of the expressions of the total charge. Let us consider
the integral%
\begin{equation}
\mathcal{P}=\int_{0}^{\infty }dx\,e^{-x}\left[ \mathcal{I}(q,\alpha
_{0},x)-e^{x}/q\right] ,  \label{IntP}
\end{equation}%
where $\mathcal{I}(q,\alpha _{0},x)$ is given by (\ref{series}) with $\alpha
=\alpha _{0}$. From the formula (\ref{Ic1}) it follows that the integral is
convergent. We evaluate it in two different ways. Firstly, we use the
expression (\ref{series}). Inserting into (\ref{IntP}), we note that the
separate integrals with the first and second terms in the square brackets
diverge. In order to have the right for separate integrations, we write the
integral as $\lim_{\lambda \rightarrow 1}\int_{0}^{\infty }dx\,e^{-\lambda x}%
\left[ \mathcal{\cdots }\right] $. For $\lambda >1$ both the integrals
converge separately. By using the standard result for the integral with the
modified Bessel function \cite{Prud86}, after the summation over $j$ and the
limiting transition $\lambda \rightarrow 1$ we get%
\begin{equation}
\mathcal{P}=\frac{1-q^{2}}{12q}+\alpha _{0}\left( q\alpha _{0}-1\right) .
\label{P2}
\end{equation}

In the second approach for the evaluation of the integral (\ref{IntP}) we
use the representation (\ref{Ic1}). After the elementary integration over $x$
one finds%
\begin{equation}
\mathcal{P}=\frac{1}{q}\sum_{l=1}^{p}\frac{(-1)^{l}}{s_{l}^{2}}\cos (2\pi
l(\alpha _{0}-1/2q))-\frac{1}{\pi }\int_{0}^{\infty }dy\,\frac{f(q,\alpha
_{0},2y)/\cosh ^{2}y}{\cosh (2qy)-\cos (q\pi )}.  \label{P3}
\end{equation}%
From (\ref{P2}) and (\ref{P3}) it follows that%
\begin{equation}
\sum_{l=1}^{p}\frac{(-1)^{l}}{s_{l}^{2}}\cos (2\pi l(\alpha _{0}-1/2q))-%
\frac{q}{\pi }\int_{0}^{\infty }dy\,\frac{f(q,\alpha _{0},2y)/\cosh ^{2}y}{%
\cosh (2qy)-\cos (q\pi )}=\frac{1-q^{2}}{12}+q\alpha _{0}\left( q\alpha
_{0}-1\right) .  \label{Rel4}
\end{equation}%
In the special case $\alpha _{0}=0$ this gives:%
\begin{equation}
\sum_{l=1}^{p}\frac{(-1)^{l}}{s_{l}^{2}}c_{l}+\frac{2q}{\pi }\cos (q\pi
/2)\int_{0}^{\infty }dy\,\frac{\sinh (qy)\sinh y\cosh ^{-2}y}{\cosh
(2qy)-\cos (q\pi )}=\frac{1-q^{2}}{12}.  \label{Rel4alf0}
\end{equation}%
As another consequence of (\ref{Rel4}) one has%
\begin{equation}
\sum_{l=1}^{p}\frac{(-1)^{l}}{s_{l}}\sin (2\pi l\alpha _{0})-\frac{q}{\pi }%
\int_{0}^{\infty }dy\,\frac{f_{2}(q,\alpha _{0},y)/\cosh ^{2}y}{\cosh
(2qy)-\cos (q\pi )}=-q\alpha _{0}.  \label{Rel4b}
\end{equation}%
Other relations are obtained from (\ref{Rel4}) by differentation with
respect to $\alpha _{0}$.

\end{document}